\documentclass{aa}
\usepackage{txfonts}
\usepackage{natbib}
\bibpunct{(}{)}{;}{a}{}{,} % to follow the A&A style
\usepackage{graphicx}
\usepackage{color}
\usepackage{hyperref}
\usepackage{breakurl}
\usepackage{ulem}

\begin{document}

\title{Galaxy mass, cluster-centric distance and secular evolution: their role in the evolution of galaxies in clusters in the last 10 Gyr}
%\subtitle{Subtitle}
\author{A. Raichoor  \and S. Andreon}
\institute{INAF -- Osservatorio Astronomico di Brera, via Brera 28, 20121 Milan, Italy\\
e-mails: 
\texttt{[\href{mailto:anand.raichoor@brera.inaf.it}{anand.raichoor},
\href{mailto:stefano.andreon@brera.inaf.it}{stefano.andreon}]@brera.inaf.it}}

\date{Accepted ... . Received ...}
%\pagerange{\pageref{firstpage}--\pageref{lastpage}} \pubyear{2002}
%\maketitle
%\label{firstpage}

\abstract
{Galaxy mass and environment are known to play a key role in galaxy evolution: studying galaxy colors as a function of redshift, galaxy mass, and environment offers a powerful diagnosis to disentangle the role of each.}
{We study the simultaneous dependence of the fraction of blue galaxies $f_{blue}$ on secular evolution, environment, and galaxy mass with a well-controlled cluster sample.
We are thus able to study the evolution and respective role of the cessation of star formation history (SFH) in clusters caused by galaxy mass ("mass quenching") or by environment ("environmental quenching").}
{We defined an homogenous X-ray selected cluster sample (25 clusters with $0 < z < 1$ and one cluster at $z \sim 2.2$), having similar masses and well-defined sizes.
Using multicolor photometry and a large spectroscopic sample to calibrate photometric redshifts, we carefully estimated $f_{blue}$ for each cluster at different galaxy mass and cluster-centric distance bins.
We then fitted the dependence of $f_{blue}$ on redshift ($z$), environment ($r/r_{200}$) and galaxy mass ($M$) with a simple model.}
{$f_{blue}$ increases with cluster-centric distance with a slope $\alpha = 1.2_{-0.3}^{+0.4}$, decreases with galaxy mass with a slope  $\beta = -3.8_{-0.5}^{+0.6}$, and increases with redshift with a slope $\gamma = 3.2_{-0.5}^{+0.7}$.
The data also require for the first time a differential evolution with galaxy mass of $f_{blue}$ with redshift, with lower mass galaxies evolving slower by a factor $\zeta = -4.1_{-0.9}^{+1.1}$.}
{Our study shows that the processes responsible for the cessation of star formation in clusters are effective at all epochs ($z \lesssim 2.2$), and more effective in denser environments and for more massive galaxies. 
We found that the mass and environmental quenchings are separable, that environmental quenching does not change with epoch, and that mass quenching is a dynamical process, i.e. its evolutionary rate is mass-dependent.
Our study extends the \textit{downsizing}-like scenario, where the most massive galaxies have their properties set at a very high redshift, to the cluster environment and all galaxies.
It illustrates the need to disentangle galaxy mass and cluster-centric distance to properly estimate the behavior of $f_{blue}$ in clusters.}

\titlerunning{Evolution of galaxies in clusters in the last 10 Gyr}
\authorrunning{A. Raichoor \& S. Andreon}
\maketitle

%@@@@@@@@@@@@@@@@@@@@@@@@@@@@@@@@@@@@@@@@@@@
% INTRODUCTION
%@@@@@@@@@@@@@@@@@@@@@@@@@@@@@@@@@@@@@@@@@@@

\section{Introduction \label{sec:intro}}

% INTRO FOR MASS/ENVIRONMENT
It is well-established that galaxy mass and environment are key parameters in shaping galaxy properties \citep[e.g.,][]{butcher78,dressler80,de-propris03}.
At least in the local Universe, both change galaxy properties in a similar direction, for instance making them preferentially redder \citep[e.g.,][]{visvanathan77,butcher78} or of early-type morphology \citep[e.g.,][]{blanton03a,dressler80}.
Furthermore, there is a correlation between galaxy mass and environment, because denser environments tend to be inhabited by more massive galaxies \citep[e.g.,][]{hogg03,baldry06}.
In the local Universe, the Sloan Digital Sky Survey \citep[SDSS,][]{york00} provided data from which one could quantify the role of each to an unprecedented precision \citep[e.g.,][]{kauffmann04,peng10}.

% SECULAR EVOLUTION/MASS/ENVIRONMENT
However, when considering higher redshifts ($z \sim 1$-2), the analysis is complicated -- and additionally there are less data available -- by the disentangling of the precise role of secular galaxy evolution, of galaxy mass, and of environment on galaxy properties.
Therefore, any analysis of galaxy evolution at $z \sim 1$-2 needs to \textit{simultaneously} control for galaxy secular evolution, galaxy mass, and environment.
Indeed, any bias in one of those three terms may confuse the interpretation of the obtained result.
For instance, a blueing related to a younger age at higher redshift may be mistaken for a blueing induced by a sampling biased preferentially toward lower density environments, which are richer in blue galaxies.

% GALAXY MASS CONTROL
Controlling for galaxy mass requires working with galaxy mass-selected samples, which presents the advantage of a better control of the sample.
Indeed, galaxy mass has on average a much more regular evolution (increasing with decreasing redshift) than galaxy rest-frame optical luminosity, which can temporarily increase with starbust, and then decrease.
For example, a sample selected on rest-frame optical luminosity is very likely to be biased toward those temporary low-mass starbust galaxies at high redshifts, thus introducing a spurious increase in the fraction of blue galaxies that disappears if a mass-selected sample is used \citep[e.g.,][]{de-propris03}.
A more subtle question is the mass evolution of star-forming galaxies that will significantly increase their stellar mass with time.
For example, an average star formation rate of 2 $M_{\sun}$ yr$^{-1}$ (resp. 5 $M_{\sun}$ yr$^{-1}$) for 5 Gyr (resp. 2 Gyr) is sufficient to build $10^{10} M_{\sun}$.
Therefore, selecting galaxies by stellar mass measured at the galaxy redshift may introduce a net inflow with decreasing redshift of galaxies that rise above the mass threshold adopted for the study.
This uncontrolled inflow with decreasing redshift of galaxies in the selected sample leaves the ambiguity between a real evolution of massive blue galaxies and a selection effect.
To remove this ambiguity, one needs to control for mass not at the redshift of observation, but evolved at $z = 0$ to select at high redshift the likely ancestors of present-day galaxies of a given mass \citep[e.g.,][]{andreon08a,raichoor12}.

% GALAXY AGING WITH REDSHIFT
Controlling for secular evolution requires acknowledging that the galaxy rest-frame color evolves with time.
Indeed, galaxies observed at higher redshifts will on average have bluer rest-frame colors than their local counterparts, because of their mean younger age and of the higher mean star formation activity of the Universe at higher redshifts \citep[e.g.,][]{lilly96,madau98}.
Using a non-evolving rest-frame color will classify an increasing number (with increasing redshift) of red-sequence galaxies as blue galaxies, thus possibly introducing a spurious evolutionary trend.
One can take this into account by using evolving rest-frame colors to characterize galaxy properties at different redshifts, as proposed by \citet{andreon06} and now adopted in several works \citep[e.g.,][]{haines09,peng10,raichoor12}.

% ENVIRONMENT CONTROL
Controlling for environment raises the question of how to measure environment.
Available data often oblige us to use a proxy for an environment estimate based on samples with either photometric redshift or that are rest-frame $B$-band luminosity-selected.
However, at $z \sim 1$-2, these proxies may be either prone to large uncertainties or potentially biased toward (temporarily) blue overdensities.
Again, this may introduce spurious trends or smooth existing ones \citep[for instance, see discussion in \S3 of][]{quadri12}.
In this regard, clusters of galaxies are ideal laboratories.
On the one hand, their cores unambiguously represent the densest environment in the Universe at each epoch, where environmental processes consequently are the most effective.
On the other hand, they provide, at fixed redshift, handy samples of galaxies observed in similar conditions, with little contamination.
Nevertheless, when using cluster-centric distance as the environment's measurement in clusters, one should scale it with the cluster size (e.g. $r_{200}$): if not, one risks to assign the same environmental measurement to different physical environments, thus mixing age-related trend (redshift dependence) with environmental trend.
The effect is of paramount importance because gradients in cluster populations are usually important.
For instance, if one probes a cluster region within a fixed radius (e.g. 1 Mpc), one will probe only central regions of large clusters but outer regions for small clusters, which possibly introduces an artificial dependence with respect to the cluster size \citep[e.g.][]{margoniner01}.
Moreover, even if challenging at $z \sim 1$-2, a robust cluster size estimation is needed to prevent any smoothing of possible environmental trends when co-adding data \citep[see for instance][and the use of the approximate cluster size estimate $B_{GC}$]{loh08}.

% CLUSTER MASS
Additionally, if one uses galaxy clusters to study galaxy evolution, the cluster selection function should be as independent as possible of the (studied) galaxy population.
As an example, by selecting preferentially clusters rich in blue galaxies at higher redshift \citep[because they are easier to detect, as in the early 80's,][]{kron95} or 
richer in red galaxies \citep[because clusters are detected thanks to the red sequence, as in ][with the RCS]{loh08} likely introduces a selection effect when estimating the fraction of blue galaxies \citep{andreon06}.
It is preferable to select clusters by their X-ray emission, because at a given X-ray emission ($L_X$ or $T_X$) clusters rich in blue galaxies are not favored/disfavored at a given cluster mass \citep{andreon99}.
Furthermore, it is well-known that cluster galaxy properties depend on the cluster richness \citep[e.g.,][]{oemler74,dressler80,dressler97,hansen09}: a way to control this dependence is to select clusters with similar masses.

% THIS WORK
In this paper, we aim at disentangling the role of galaxy secular evolution, galaxy mass, and environment on galaxy evolution by analyzing a cluster sample spanning a wide redshift baseline.
An observationally economic but still physically efficient proxy for measuring the star formation activity is galaxy color \citep[e.g.,][]{butcher84,williams09,peng10}.
We  classified galaxies into two broad categories (blue or red) and defined the fraction of blue galaxies $f_{blue}$ at a given galaxy mass $M$ and scaled cluster-centric distance $r/r_{200}$, in the vein of the  pioneering work of \citet{butcher84}.
In the last decades, numerous works studied $f_{blue}$ in clusters \citep[e.g.,][to name a few]{de-propris03,andreon06,loh08,haines09} but none of them studied
the dependence of $f_{blue}$ on redshift, galaxy mass, and cluster-centric distance at the same time.
Analyzing simultaneously the dependence of $f_{blue}$ on those three parameters allows one to put strong constraints on the various processes that are responsible for the cessation of star formation activity ("quenching"). 

% THIS WORK: METHOD
To achieve this, we considered the aforementioned remarks to our best ability.
We used a sizable X-ray selected cluster sample (25 clusters with $0 < z < 1$ and the JKCS\,041 cluster at $z \sim 2.2$) with well-defined $r_{200}$ and similar cluster masses, thus minimizing the correlation between $f_{blue}$ and our cluster selection function.
We then estimated $f_{blue}$ as a function of galaxy mass $M$ and scaled cluster-centric distance $r/r_{200}$, defining blue/red galaxies considering also the stellar evolution of galaxies with time.
To lead the analysis, we used a reasonable evolutionary model as a reference (exponentially declining star formation history, fixed formation redshift), and then measured any deviation from this reference.
This reference model does not aim to describe the individual behavior of each galaxy, but to cancel out overall trends such as stellar aging or stellar mass increasing with time.
Deviations from this reference model highlight the evolution and respective role of the quenching in clusters caused by galaxy mass or to environment.

% PAPER OUTLINE
The plan of this paper is as follows.
We define in Sect. \ref{sec:sample} the cluster sample on which this study relies.
Sect. \ref{sec:data} presents the data and their analysis, including our estimation of $f_{blue}$.
Our results are presented in Sect. \ref{sec:results} and are summarized and discussed in Sect. \ref{sec:conclusion}.
% PAPER CONVENTIONS
We adopt $H_0 = 70$ km s$^{-1}$ Mpc$^{-1}$,  $\Omega_m = 0.30$, and $\Omega_\Lambda = 0.70$ throughout.
All magnitudes are in the AB system, corrected for Galactic extinction using \citet{schlegel98}.
Masses are computed with a \citet{chabrier03} initial mass function and, if not stated otherwise, are defined by the mass of the gas that will eventually be turned into stars, i.e. corresponding to the integral of the star formation rate.

%@@@@@@@@@@@@@@@@@@@@@@@@@@@@@@@@@@@@@@@@@@@
% CLUSTER SAMPLE
%@@@@@@@@@@@@@@@@@@@@@@@@@@@@@@@@@@@@@@@@@@@

\section{Cluster sample \label{sec:sample}}

\begin{figure}
\includegraphics[width=\linewidth]{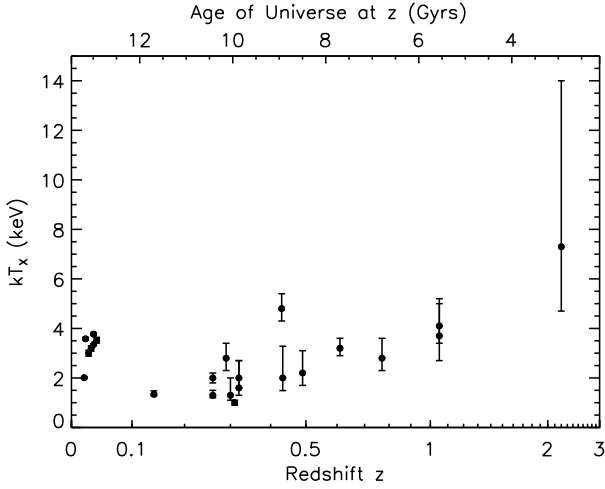}
\caption{
$T_X$ as a function of redshift for our cluster sample.
\label{fig:Tx_z}
}
\end{figure}

We describe in this section our cluster sample.
It was assembled from an X-ray selection and clusters with similar masses at different redshifts, using the cluster X-ray temperature as mass proxy.
The X-ray selection ensures that our cluster sample selection is unbiased by $f_{blue}$, because the probability of inclusion of a cluster in the sample is independent of $f_{blue}$ at a given cluster mass.
At intermediate redshift, we relied on the X-ray selected cluster sample from the first five deg$^2$ of the XMM Large-Scale Structure survey \citep[XMMLSS,][]{pierre04}.
At low redshift, we extracted a sample from the HIghest X-ray FLUx Galaxy Cluster Sample \citep[HIFLUGCS,][]{reiprich02}, an X-ray selected cluster sample, by selecting clusters with comparable cluster masses.
We added to this sample the JKCS\,041 cluster \citep[$z \sim 2.2$,][]{andreon09,andreon11}.
Though this cluster is different from the other objects, because it is not X-ray selected, it is the highest redshift cluster with a significant X-ray emission; moreover, its mass/temperature is comparable to the rest of our sample.
Figure \ref{fig:Tx_z} displays the temperature $T_X$ of our cluster sample as a function of redshift.
We note that our clusters (except for JKCS\,041) have $kT \sim 3$ keV, typical of intermediate mass clusters.
We have $r_{200}$ for all the objects,  which is of paramount importance, because it is known that $f_{blue}$ depends on cluster-centric distance.
It has been derived from $T_X$ or $\sigma_v$ as described in Appendix \ref{app:r200}.
The properties of the clusters are listed in Table \ref{tab:cluster_sample}.

% FIGURE: Tx vs z

% TABLE: CLUSTER SAMPLE
\begin{table}
\centering
\caption{Cluster sample \label{tab:cluster_sample}}
\begin{tabular}{l l l l l}
\hline
\hline
Cluster		& $z_{spec}$ &   $kT_X$   &       $\sigma_v$              &       $r_{200}$\\
                       	&                       &       (keV)           &       (km s$^{-1}$)   &       (Mpc)\\
\hline
\multicolumn{5}{c}{SDSS - HIFLUGCS}\\
\hline
MKW4    &       0.020   &       2.01$_{-0.04}^{+0.04}$  &       -       &       0.91\\
A1367   &       0.022   &       3.58$_{-0.06}^{+0.06}$  &       -       &       1.31\\
MKW8    &       0.027   &       3.00$_{-0.12}^{+0.12}$  &       -       &       1.17\\
A2634   &       0.031   &       3.19$_{-0.11}^{+0.11}$  &       -       &       1.21\\
A2052   &       0.035   &       3.35$_{-0.02}^{+0.02}$  &       -       &       1.25\\
A2063   &       0.035   &       3.77$_{-0.06}^{+0.06}$  &       -       &       1.34\\
A2657   &       0.040   &       3.52$_{-0.11}^{+0.12}$  &       -       &       1.28\\
\hline
\multicolumn{5}{c}{CFHTLS W1 - XMMLSS}\\
\hline
XLSSC\,041      &       0.14    &       1.3$_{-0.1}^{+0.1}$     &       -       &       0.66\\
XLSSC\,044      &       0.26    &       1.3$_{-0.1}^{+0.2}$     &       -       &       0.61\\
XLSSC\,025      &       0.26    &       2.0$_{-0.2}^{+0.2}$     &       -       &       0.80\\
XLSSC\,027      &       0.29    &       2.8$_{-0.5}^{+0.6}$     &       -       &       0.98\\
XLSSC\,008      &       0.30    &       1.3$_{-0.2}^{+0.7}$     &       -       &       0.60\\
XLSSC\,013      &       0.31    &       1.0$_{-0.1}^{+0.1}$     &       -       &       0.51\\
XLSSC\,040      &       0.32    &       1.6$_{-0.3}^{+1.1}$     &       -       &       0.67\\
XLSSC\,018      &       0.32    &       2.0$_{-0.4}^{+0.7}$     &       -       &       0.78\\
XLSSC\,016      &       0.33    &       -       &       703$_{-266}^{+266}$     &       1.25\\
XLSSC\,014      &       0.34    &       -       &       416$_{-246}^{+246}$     &       0.73\\
XLSSC\,006      &       0.43    &       4.8$_{-0.5}^{+0.6}$     &       -       &       1.26\\
XLSSC\,012      &       0.43    &       2.0$_{-0.5}^{+1.3}$     &       -       &       0.73\\
XLSSC\,049      &       0.49    &       2.2$_{-0.5}^{+0.9}$     &       -       &       0.75\\
XLSSC\,007      &       0.56    &       -       &       323$_{-191}^{+178}$     &       0.50\\
XLSSC\,001      &       0.61    &       3.2$_{-0.3}^{+0.4}$     &       -       &       0.88\\
XLSSC\,002      &       0.77    &       2.8$_{-0.5}^{+0.8}$     &       -       &       0.74\\
XLSSC\,029      &       1.05    &       4.1$_{-0.7}^{+0.9}$     &       -       &       0.79\\
XLSSC\,005      &       1.05    &       3.7$_{-1.0}^{+1.5}$     &       -       &       0.74\\
\hline
JKCS\,041       & $\sim 2.2$    &       7.3$_{-2.6}^{+6.7}$             &       -       &       0.76\\
\hline
\end{tabular}
\tablefoot{
$T_X$ are taken from \citet{hudson10} for the SDSS/HIFLUGCS subsample, \citet{pacaud07} for the CFHTLS W1/XMMLSS subsample and \citet{andreon11b} for JKCS\,041.
$\sigma_v$ are from \citet{willis05}. $r_{200}$ are estimated from $T_X$, except for for clusters XLSSC\,007, XLLSC\,014 and XLSSC\,016, which are estimated from $\sigma_v$ (cf. Appendix \ref{app:r200}).
}
\end{table}

%------------------------------------------------------------------------------------------------------------------------
% XMMLSS SAMPLE
%------------------------------------------------------------------------------------------------------------------------
\subsection{XMMLSS sample ($0.14 \le z \le 1.05$)}

\begin{figure*}
\includegraphics[width=\linewidth]{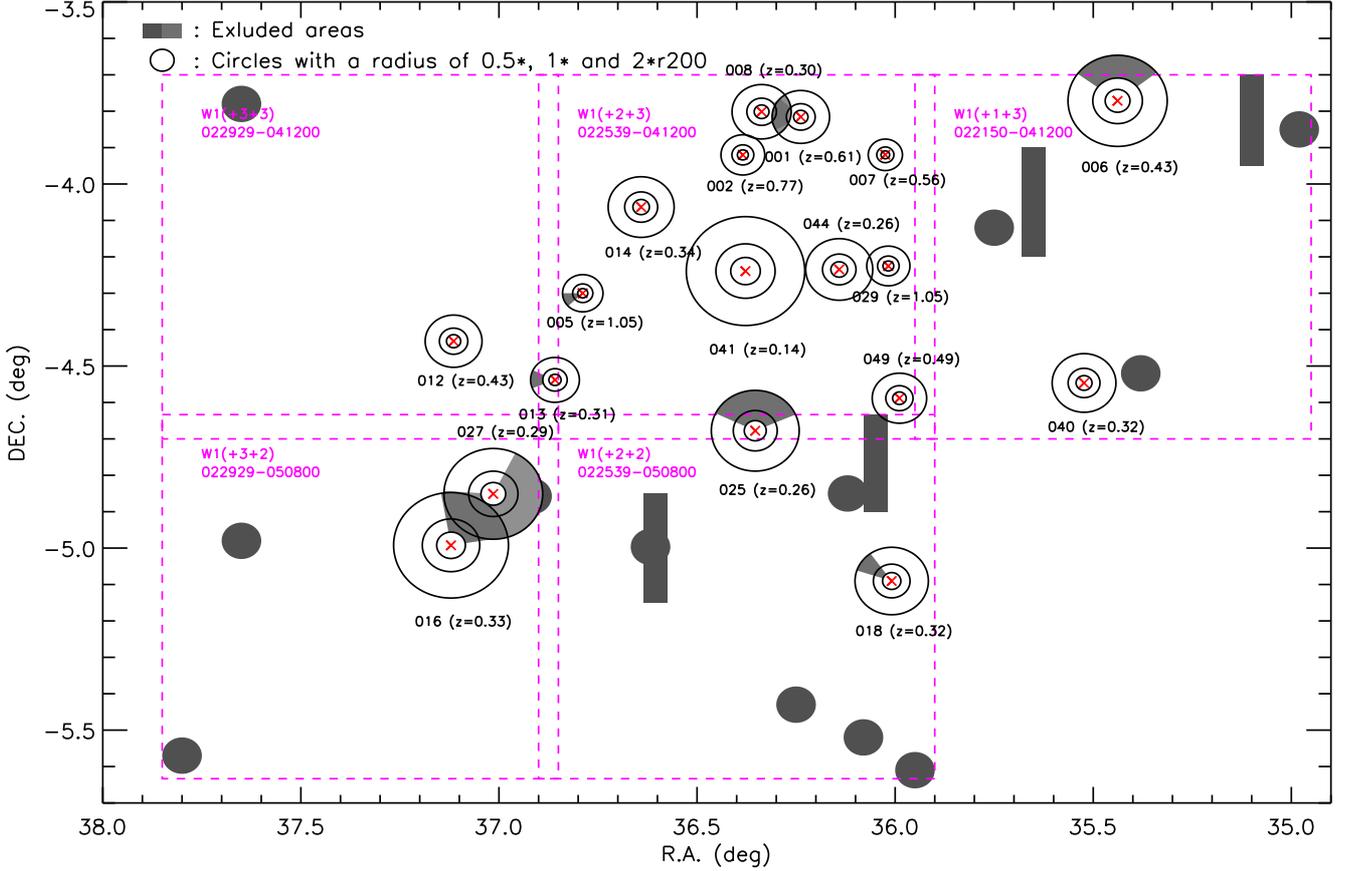}
\caption{
CFHTLS W1 cluster subsample: spatial location of the 18 clusters of our CFHTLS W1 subsample (red crosses).
The magenta dashed lines represents CFHTLS W1 fields.
For each cluster, blue circles have radii of $0.5 \times$, $1 \times$ and $2 \times r_{200}$.
Gray areas represent masked regions for $f_{blue}$ estimation and background estimation, due to image corruption (rectangles) or potential clusters at similar redshifts as those of our cluster subsample for each field (disks and sectors).
}
\label{fig:CFHTLS_W1_clusters}
\end{figure*}

The main body of our cluster sample comes from the XMMLSS.
This survey is located in the W1 area of the Canada-France-Hawaii Telescope Legacy Survey (CFHTLS), which provides $u^*g'r'i'z'$-band data.
More precisely, we selected all clusters from the first five deg$^2$ \citep{valtchanov04,willis05,pacaud07}, which are in the CFHTLS W1 field  (see Figure \ref{fig:CFHTLS_W1_clusters}).
We removed clusters with lower-than-average data quality:
XLSSC\,011 and XLSSC\,021 (their galaxies saturate CFHTLS W1 data because they are too nearby),
XLSSC\,004 (no $T_X$ or $\sigma_v$ measurements, so we cannot estimate $r_{200}$),
XLSSC\,017 and XLSSC\,020 (located where CFHTLS W1 data are corrupted),
XLSSC\,022 (blended with clusters at similar redshifts).
We finally assembled 18 clusters, 15 of which having $r_{200}$ estimated from $T_X$ measurements with the \textit{XMM} telescope \citep{pacaud07},  3 (XLSSC\,007, XLSSC\,014, and XLSSC\,016) from $\sigma_v$ measurements \citep{willis05}.

% FIGURE: CFHTLS CLUSTERS

%------------------------------------------------------------------------------------------------------------------------
% HIFLUGCS SAMPLE
%------------------------------------------------------------------------------------------------------------------------
\subsection{HIFLUGCS sample ($0.02 \le z \le 0.05$)}

\begin{figure*}
\includegraphics[width=\linewidth]{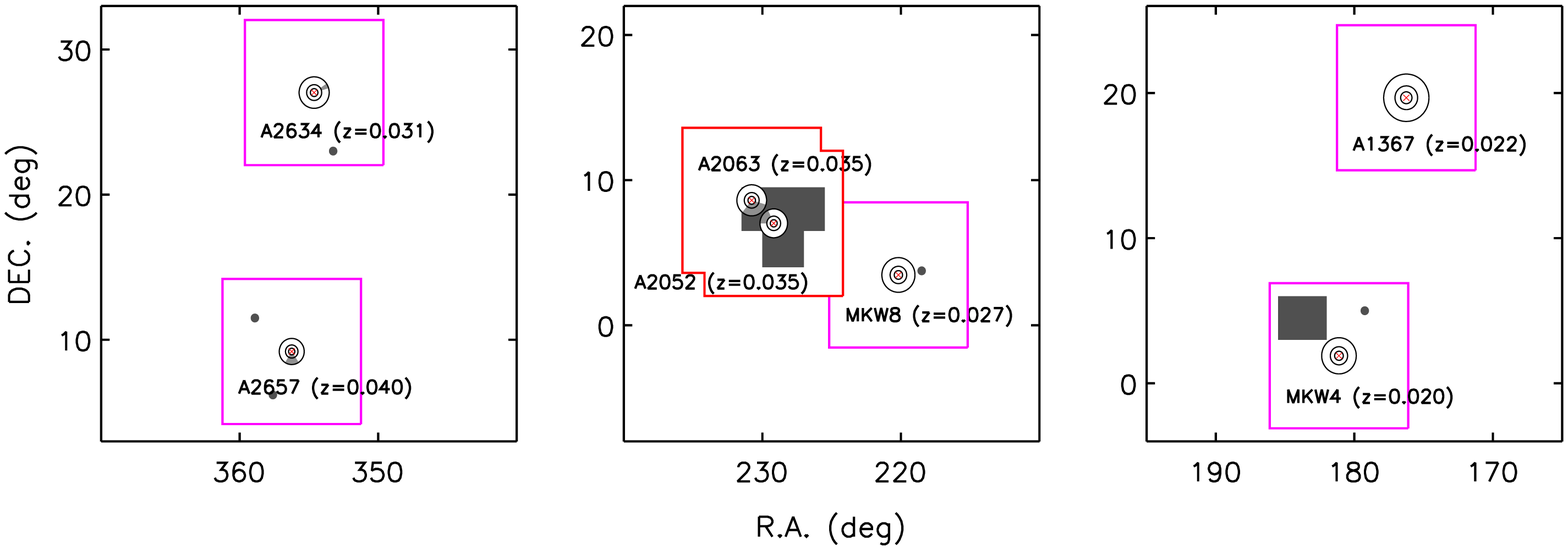}
\caption{
SDSS cluster subsample: spatial location of the seven clusters of our SDSS subsample (red crosses).
Magenta/red solid lines represent our SDSS field used for background estimation for each cluster (see Sect. \ref{sec:SDSS}).
For each cluster, blue circles have radii of $0.5 \times$, $1 \times$ and $2 \times r_{200}$.
Gray areas represent masked regions for $f_{blue}$ estimation and background estimation, due to large-scale structures or potential clusters at similar redshifts as those of our cluster subsample for each field.
}
\label{fig:SDSS_clusters}
\end{figure*}

We completed our cluster sample with a low-redshift cluster sample, with similar masses ($T_X$).
We started from the X-ray selected and X-ray flux-limited HIFLUGCS, all of which have \textit{Chandra} temperature measurements \citep{hudson10}.
Out of the 64 clusters from the HIFLUGCS sample, we selected those with $z_{spec} \ge 0.02$ (to avoid shredding), $kT_X(keV) \le 3.8$ (to match cluster masses) and those falling in the SDSS field, thus obtaining seven low-redshift clusters.
Their spatial positions are displayed in Figure \ref{fig:SDSS_clusters}.

% FIGURE: SDSS CLUSTERS

%@@@@@@@@@@@@@@@@@@@@@@@@@@@@@@@@@@@@@@@@@@@
% DATA AND ANALYSIS
%@@@@@@@@@@@@@@@@@@@@@@@@@@@@@@@@@@@@@@@@@@@

\section{Data and analysis \label{sec:data}}

To measure $f_{blue}$, we adopted a conservative approach which may return large error bars, but yields unbiased values.
For instance, to take into account the background (resp. stars), we adopted a minimum removal, that statistically accounts for the remaining background (resp. stars).

% METHOD
To this aim, we needed to
a) identify and remove stars;
b) correct the underestimate of photometric errors listed in the original catalog, if any;
c) correct for a (minor) residual photometric offset;
d) estimate redshift probability distribution functions $p(z)$;
e) mask potential contamination from other clusters at similar redshift and remove background sources;
f)  make sure that the different subfields have consistent colors, to prevent any systematic offset in the estimated photometric redshifts.

We describe in  Sect. \ref{sec:CFHTLS} the data used for the CFHTLS W1/XMMLSS subsample, along with their analysis.
The procedure for the SDSS/HIFLUGCS subsample is similar, and is summarized in Sect. \ref{sec:SDSS}.
The analysis of JKCS\,041 is described in \citet{raichoor12}.
We describe our estimation of $f_{blue}$ in Sect. \ref{sec:boe}.

We emphasize that we adopted a consistent method for all three cluster subsamples.

%------------------------------------------------------------------------------------------------------------------------
% CFHTLS
%------------------------------------------------------------------------------------------------------------------------
\subsection{CFHTLS W1 \label{sec:CFHTLS}}

%******************************
% DATA
%******************************
\subsubsection{Data}

% CFHTLS W1	
For the CFHTLS W1/XMMLSS subsample, we used the CFHTLS-T0006 data release of merged source catalogs ($u^*g'r'i'z'$ bands, catalogs are available at the CFHT Science Data Archive site \footnote{\url{http://www1.cadc-ccda.hia-iha.nrc-cnrc.gc.ca/cfht/T0006.html}}).
% VVDS
The VIMOS VLT Deep Survey project \citep[VVDS,][]{le-fevre05} gives spectroscopic redshifts, $z_{spec}$, of several thousands of objects in the same area.
Restricting the sample to  $z' \le 22.5$ (which corresponds to the faintest $z'$-band magnitude we work with), valid photometry and a secured $z_{spec}$ (flag=3,4),  yields a spectroscopic sample of 1652 galaxies ($z_{spec} \lesssim 1$) and 332 stars ($z_{spec} = 0$).

%******************************
% STAR REMOVAL
%******************************
\subsubsection{Star removal}

Stars spectroscopically identified by the VVDS cover less than 1 deg$^2$.
Therefore, we chose the following method to homogeneously remove stars in our sample.
We adopted a conservative approach for star removal by removing objects qualified as stars by \citet{coupon09} according to their color and size that also have a photometric redshift incompatible with that of the cluster.
Indeed, when cross-matched with objects that have a VVDS spectrum,  3\% of galaxies are wrongly identified as stars in \citet{coupon09}.
Although this is a low percentage, we additionally reduced the false identification rate by re-including objects classified as stars in \citet{coupon09}, but verified that $\int_{z_{cl}-0.3 \times z_{cl}}^{z_{cl}+0.3 \times z_{cl}} p(z)dz \ge 0.9$, where $z_{cl}$ is the cluster redshift and $p(z)$ is the photometric redshift probability distribution function output by \textsc{Eazy} (see Sect. \ref{sec:zphot}).

Stars not identified as such in this phase (12\% of our VVDS stars) were removed at a later stage, during the photometric redshift selection and the  background statistical subtraction phase.

%******************************
% SEXTRACTOR PHOTOMETRIC ERRORS CORRECTION
%******************************
\subsubsection{Photometric error correction}

We corrected the SExtractor \citep{bertin96} flux errors underestimation caused by the correlation of adjacent pixels during image  resampling \citep[e.g.][]{casertano00,andreon01,raichoor12} as in \citet{raichoor12}, and found a factor of 1.5, in agreement with previous studies \citep[e.g.][]{ilbert06a,coupon09,raichoor12}.

%******************************
% PHOTOMETRIC CALIBRATION
%******************************
\subsubsection{Photometric calibration \label{sec:photcalib}}

To prevent any systematic offset in colors between fields, which can lead to systematic offsets in the $p(z)$ functions, we matched the color-color diagrams of stars falling in different fields.
We selected a reference field (022539-041200), and bright stars ($i' < 21$, $r2 < 3$ pixels, $S/N>20$ in all bands, \texttt{terapix\_flag} $\le 1$).
We adopted a $\chi^2$ minimization with 3$\sigma$-clipping of stellar sequences in color-color diagrams, holding fixed all parameters but the intercept, which was frozen at the best-fit value of the reference field.
This relative photometric calibration (see Table \ref{tab:mag_correction}) ensures that one measures unbiased $p(z)$ all across our CFHTLS W1 fields.

% TABLE: PHOTOMETRIC SHIFTS APPLY FOR THE CALIBRATION
\begin{table}
\centering
\caption{Photometric correction $\delta$ applied to CFHTLS W1 subfields ($m_{new} = m_{old} + \delta$) \label{tab:mag_correction}}
\begin{tabular}{c c c c c}
\hline
\hline
Field				&	$u^*$	&	$g'$		&	$i'$		&	$z'$\\
\hline
022929-041200	&	-0.092	&	0.004	&	0.001	&	0.006	\\
022929-050800	&	0.015	&	0.011	&	0.040	&	0.015	\\
022539-050800	&	-0.100	&	-0.013	&	0.007	&	-0.021	\\
022150-041200	&	0.009	&	0.024	&	0.039	&	0.037	\\
\hline
\end{tabular}
\tablefoot{We adopted 022539-041200 field and $r'$-band magnitudes as a reference.}
\end{table}

%******************************
% PHOTOMETRIC REDSHIFTS
%******************************
\subsubsection{Redshift probability distributions $p(z)$ \label{sec:zphot}}

To remove galaxies that are very probably in front of or behind the cluster (cf. \S \ref{sec:bkg_rmv}), we estimated the full photometric redshift probability distribution function $p(z)$ for each galaxy, using \textsc{Eazy} \citep{brammer08} with default settings and an $r$-band magnitude prior (see Appendix \ref{app:r_prior}).
Systematic offsets in the photometric calibration between data and models \citep[e.g.,][]{brodwin06,brammer08} were estimated with the same approach as in \citet{raichoor12}: the found offsets are listed in Table \ref{tab:mag_offsets}.
These small shifts, comparable with previous works \citep{ilbert06a,ilbert09,coupon09,barro11,raichoor12}, were applied only during the $p(z)$ estimation.

% TABLE: PHOTOMETRIC OFFSETS FOR ZPHOT
\begin{table}
\centering
\caption{Systematic offsets $m_{fit}-m_{meas}$ between measured and best-fit model magnitudes for photometric redshift estimation \label{tab:mag_offsets}}
\begin{tabular}{l l l l l l}
\hline
\hline
Survey		&	$u$	&	$g$	&	$r$	&	$i$	&	$z$\\
\hline
CFHTLS W1	&	-0.07	&	\phantom{-}0.05	&	0.01	&	-0.03	&	\phantom{-}0.01\\
SDSS		&	-0.01&	-0.01	&	0.02	&	\phantom{-}0.02	&	-0.05\\
\hline
\end{tabular}
\end{table}

%******************************
% BACKGROUND REMOVAL
%******************************
\subsubsection{Background removal \label{sec:bkg_rmv}}

We need to account for galaxies along the cluster line of sight, which we generally call background.
We again took a conservative approach by only removing galaxies that are very probably in front of or behind the cluster, and by statistically accounting for the residual background in a second step.
	
\begin{figure}
\includegraphics[width=\linewidth]{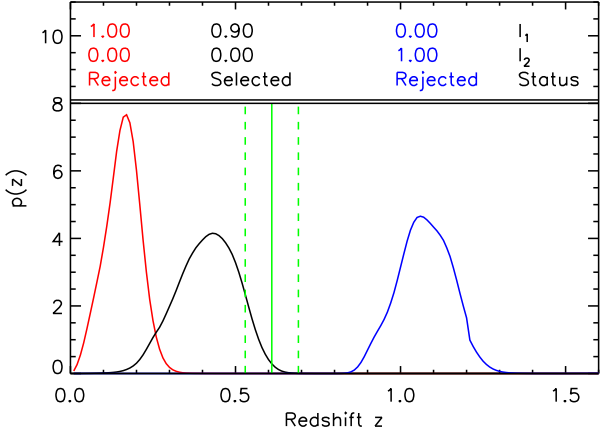}
\caption{
Illustration of the selection criterion applied in Eqs.(\ref{eq:zphot_crit1}) and (\ref{eq:zphot_crit2}) with $z_{cl} = 0.61$: we display $p(z)$ for three galaxies.
For each galaxy, we report above the value for $I_1$ (resp. $I_2$), the integral in Eq.(\ref{eq:zphot_crit1}) (resp. Eq.(\ref{eq:zphot_crit2})), and whether the galaxy is selected or not.
Vertical green solid and dashed lines represent $z_{cl}$ and $z_{cl} \pm 0.05 \times (1+z_{cl})$, respectively.}
\label{fig:selcrit}
\end{figure}

First, we performed a photometric redshift selection by removing galaxies that are at $z<z_{cl} - 0.05 \times (1+z_{cl})$ or $z > z_{cl}+ 0.05 \times (1+z_{cl})$ at $\ge 99$ \% confidence.
This selection has been calibrated on our spectroscopic VVDS sample and
was accomplished by keeping objects with
\begin{equation}
\int_0^{z_{cl}-0.05\times(1+z_{cl})} p(z)dz \le 0.99
\label{eq:zphot_crit1}
\end{equation}
\begin{equation}
\qquad \qquad \qquad \textnormal{and} \quad \int_{{z_{cl}+0.05\times(1+z_{cl}})}^{+ \infty} p(z)dz \le 0.99.
\label{eq:zphot_crit2}
\end{equation}
Figure \ref{fig:selcrit} illustrates such a selection criterion for three galaxies.

% FIGURE: SELECTION CRITERION

\begin{figure}
\includegraphics[width=\linewidth]{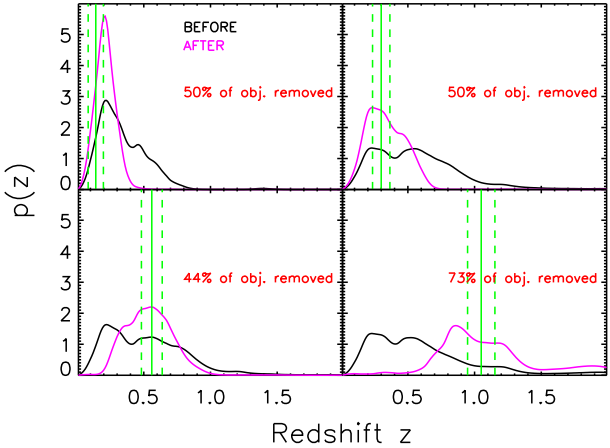}
\caption{
Cumulative $p(z)$ function for all objects in background area for four clusters, before (black curves) and after (magenta curves) applying the selection described by Eqs.(\ref{eq:zphot_crit1}) \& (\ref{eq:zphot_crit2}).
Vertical green solid and dashed lines represent $z_{cl}$ and $z_{cl} \pm 0.05 \times (1+z_{cl})$, respectively.
}
\label{fig:bkg_rmv_eff}
\end{figure}

This selection is conservative, because it keeps galaxies in the sample that have a low probability to belong to the cluster.
Eqs.(\ref{eq:zphot_crit1}) and (\ref{eq:zphot_crit2}) wrongly remove $\lesssim 1\%$ of galaxies from our spectroscopic VVDS sample (13/1652).
To illustrate the efficiency of this background removal, we display in Figure \ref{fig:bkg_rmv_eff} the cumulative $p(z)$ function for all objects in the background area for four clusters, before and after applying the selection described by Eqs.(\ref{eq:zphot_crit1}) \& (\ref{eq:zphot_crit2}).
We observe that this selection removes $\gtrsim 50\%$ of background objects.
We note that the same amount of objects are removed for cluster outer regions ($1 < r / r_{200} < 2$), but less for cluster core regions ($r/r_{200} < 1$), because the background contamination is proportionally lower towards the cluster center.

We thus observe that Eqs.(\ref{eq:zphot_crit1}) \& (\ref{eq:zphot_crit2}) fulfill our conservative approach criterion: it significantly reduces the sample, with rejecting a non-significant part ($\lesssim 1\%$) of the galaxies we are interested in.
Changing the $\{0.05,0.99\}$ values toward a more stringent selection could reduce the contamination in our sample, but at the cost of removing a more significant part of the galaxies we are interested in.
Changing the $\{0.05,0.99\}$ values toward a less stringent selection would only increase the contamination in our sample.

% FIGURE: BKG REMOV. EFFICIENCY

As a second stage of the background subtraction, we aimed to select independent control fields taken from the very same data and CFHTLS W1 field as the cluster galaxies, to guarantee homogeneity across the samples.
Furthermore, because we aimed to stack data for different clusters, we also needed independent data for the background, so that we did not overstate the quality of the control sample.

We selected for each CFHTLS W1 field a control sample in the same field in the following way.
First, we removed objects within $2 \times r_{200}$ around each cluster and within circles ($0.05^{\circ}$  radius) around potential clusters at similar redshifts (gray areas in Figure \ref{fig:CFHTLS_W1_clusters}).
The several studies on clusters in the CFHTLS W1 fields ensure that contaminating clusters are taken into account.
Then, we randomly divided the remaining objects into four subsample to reduce the effect of any potential structure, which assembled four independent control subsamples (each with a surface of $\sim$0.2 deg$^2$).
We randomly associated each cluster to a control subsample. 
When clusters in a common field had similar redshifts (always less than four in our sample), we associated them with different control subsamples in this field, thus never attributing the same background twice.

In addition, this step also subtracts any star that has not been identified as one based on its colors.

%------------------------------------------------------------------------------------------------------------------------
% SDSS
%------------------------------------------------------------------------------------------------------------------------
\subsection{SDSS \label{sec:SDSS}}

SDSS data were analyzed identically.
For the SDSS/HIFLUGCS subsample, we worked with data in $ugriz$ bands from the SDSS DR8 \citep{aihara11}.
We used the \texttt{ModelMag} for galaxy colors, and \texttt{cModelMag} and \texttt{PsfMag} as an approximate total magnitude for galaxies and stars, respectively.

Our SDSS galaxy sample is selected from the \texttt{Galaxy} view of the SDSS DR8, to which we applied a cut in magnitude (\texttt{cModelMag\_r} $< 19$), a spatial cut ($10^{\circ} \times 10^{\circ}$ around each cluster) and a criterion to reject saturated stars misclassified as galaxies (\texttt{FiberMag\_u} $\ge 22 - 0.5 \times$ \texttt{ModelMag\_u}).
By matching this catalog with the spectroscopic informations from the main galaxy sample (\texttt{petroMag\_r} $< 17.8$ in \texttt{SpecPhoto} view), we obtained $\sim$30,000 spectroscopic redshifts ($z_{spec} \lesssim 0.4$), which we used to assess the quality of our background removal.

To check that the photometry does not present any systematic offset between our fields, we used more than 10,000 spectroscopic stars from the \texttt{SpecPhoto} view located in our fields in in the same way as for the CFHTLS W1 fields (Sect. \ref{sec:photcalib}).
We found indeed no offset greater than 0.01 mag for all filters and all fields.

We estimated the $p(z)$ in in the same way as for the CFHTLS W1 fields (Sect. \ref{sec:zphot}).
We applied the photometric offsets listed in Table \ref{tab:mag_offsets}.

For background removal, we selected galaxies as in Sect. \ref{sec:bkg_rmv}.
This criterion removes $\sim$20\% of the background objects, and wrongly removes $\sim$0.05\% of the spectroscopic objects.
For each cluster, we selected as a control sample objects within a $10^{\circ} \times 10^{\circ}$ area around each cluster, excluding objects within $2 \times r_{200}$ around each cluster and within circles ($0.3^{\circ}$  radius) around potential clusters at similar redshifts.
For clusters A2052 and A2063, we defined two independent control subsamples as described in Sect. \ref{sec:bkg_rmv} and using the region within the red solid line in Figure \ref{fig:SDSS_clusters}.
The resulting control samples are $\sim$90 deg$^2$ (resp.  $\sim$50 deg$^2$) for the MKW4, A1367, MKW8, A2634, and A2657 clusters (resp. the A2052 and A2063 clusters).

%------------------------------------------------------------------------------------------------------------------------
% ESTIMATION OF FBLUE
%------------------------------------------------------------------------------------------------------------------------
\subsection{Settings for the estimation of $f_{blue}$ \label{sec:boe}}

We describe in this section our procedure for consistently measuring the fraction of blue galaxies $f_{blue}$.
The procedure described here is similar to the one followed in Section 4 of \citet{raichoor12} for JKCS\,041.

For each cluster, we considered three radial bins (annuli) at distinct cluster-centric radii, defined by $r/r_{200} \le 0.5$ , $0.5 < r/r_{200} \le 1$, and $1 < r/r_{200} \le 2$.

%******************************
% PROBED COLOR AND MASS-SELECTED SAMPLE
%******************************
\subsubsection{Probed color \label{sec:color}}

\begin{figure}
\resizebox{\hsize}{!}{\includegraphics{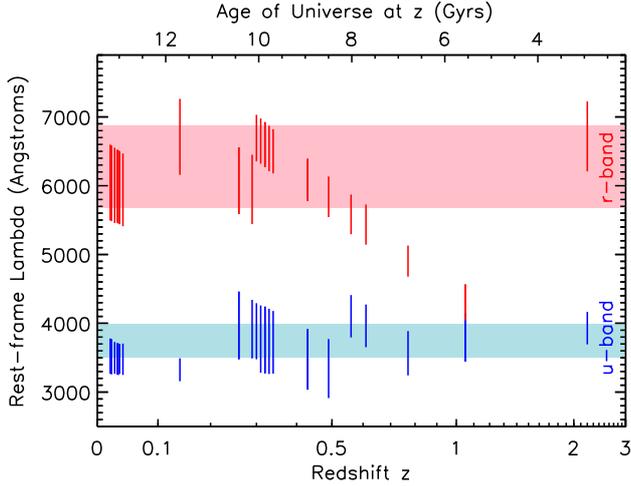}}
\caption{$(u-r)$-like rest-frame probed color: for each cluster of our sample, we plot the rest-frame probed color as a function of the redshift.
The shaded areas indicate the wavelength intervals where the filter response convolved with a typical elliptical galaxy spectrum is above half of its maximum value.
The vertical bars represent the same for the redshifts of our clusters and the corresponding chosen filters.}
\label{fig:rf_ur}
\end{figure}

To probe the same rest-frame color at all redshifts as consistently as possible, we worked in a color-magnitude diagram that closest corresponds  to the $(u-r)$ vs $r$ rest-frame, thus probing the 4000 \AA~ break.
Figure \ref{fig:rf_ur} illustrates the filters chosen at each redshift of our sample.
Because our reddest band for the CFHTLS W1 sample is the $z$ band, this implies a drift in the probed rest-frame red band when the redshift is approaching 1 ($z=0.77$ and $z=1.05$): we are here limited by the data, but still improve on previous works \citep[see for instance][]{loh08}.

% FIGURE: (U-R) REST-FRAME FILTERS

\begin{figure}
\resizebox{\hsize}{!}{\includegraphics{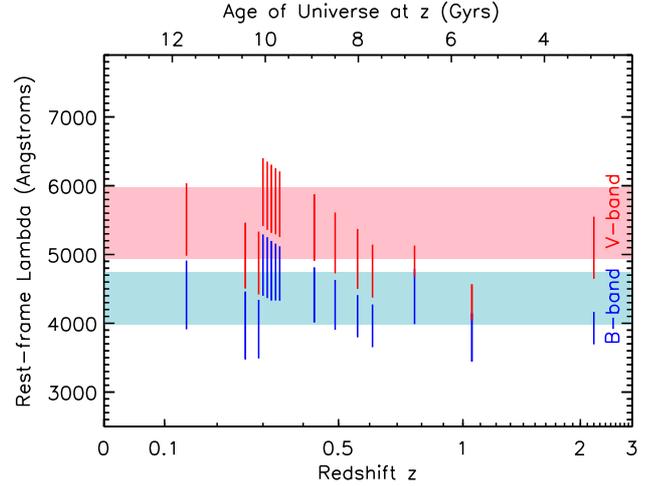}}\\
\resizebox{\hsize}{!}{\includegraphics{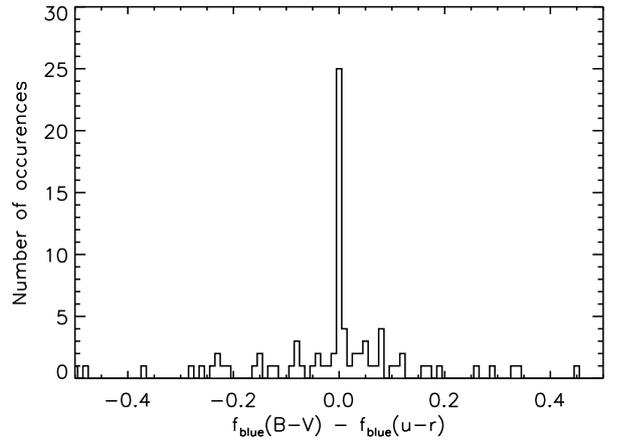}}
\caption{Dependence of $f_{blue}$ on the rest-frame probed color.
\textit{Top}: like Figure \ref{fig:rf_ur}, but for $B-V$ rest-frame color.
\textit{Bottom}: we compare the estimated $f_{blue}$ when using the rest-frame $B-V$ and $u-r$ colors
for 84 estimated values (10 clusters for which are available independent filter pairs probing the rest-frame $B-V$ and $u-r$ colors, 1 to 4 mass bins, 3 radial bins).
We stress that some individual estimates of $f_{blue}$ have large error bars (see Figure \ref{fig:BOE_indiv}): the 68\% shortest intervals for $f_{blue}(B-V)$ and $f_{blue}(u-r)$ all intersect.}
\label{fig:rf_bv}
\end{figure}

% (B-V) REST-FRAME COLOR
The top panel of Figure \ref{fig:rf_bv} illustrates another possible choice of filter pairs, matching the $(B-V)$ rest-frame color.
The bottom panel shows that $f_{blue}$ -- estimated as described below -- does not depend on the precise sampling of the filters, given that they bracket the 4000 \AA~break.
Our results are robust regarding the rest-frame color: using the $(B-V)$ color instead of $(u-r)$ leads to similar values of $f_{blue}$ (see Figure \ref{fig:rf_bv}).

% FIGURE: (B-V) REST-FRAME FILTERS + FBLUE(BV) vs FBLUE(UR)

%******************************
% MASS
%******************************
\subsubsection{Mass bin defintion \label{sec:mass}}

\begin{figure}
\includegraphics[width=\linewidth]{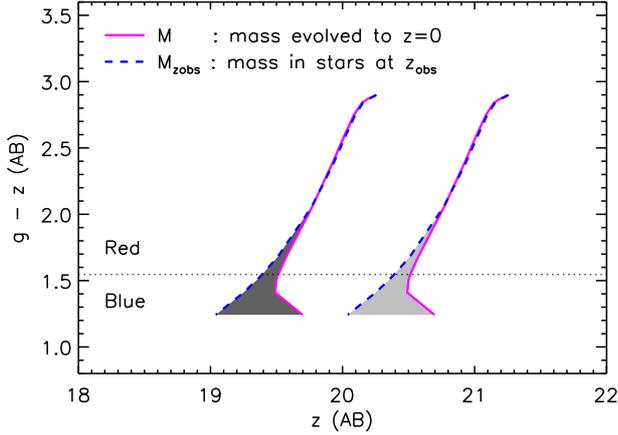}
\caption{
Illustration of a mass bin definition ($z_{obs}=0.49$, $10.73 \le \log (M/M_{\sun}) < 11.13$):
thick magenta solid lines represent the loci of constant mass $M$ used in this study (the mass evolved to $z=0$),
thick blue dashed lines the loci of constant mass $M_{zobs}$ in stars at $z_{obs}$,
and gray shaded areas represent the difference between those two definitions.
The horizontal dotted line represents the threshold used to define blue/red galaxies.
}
\label{fig:massdef}
\end{figure}

For our values of mass $M$ we refer, as in previous works, to the \cite{bruzual03} model (2007 version, CB07 hereafter) mass, and specifically to \textit{the mass of the gas that will eventually be turned into stars}, i.e. the integral of the star formation rate \citep{andreon08a,raichoor12}.
This mass definition has the advantage to compare galaxies at each epoch that will end up with the same stellar mass, which means that we include in each mass bin the same galaxies at high and low redshifts.
In the color-magnitude diagrams, the shape of these loci of constant mass differ from that which corresponds to the mass in stars at the redshift of observation ($M_{zobs}$) on the blue end (see Figure \ref{fig:massdef}): the higher the redshift, the more fainter/bluer objects are included in our definition.
For instance, a galaxy with an exponentially declining SFH $\propto \exp [-(t/\tau)]$ with $\tau = 3$ Gyr would have a mass in stars of $\sim$30/50/90\% of its final mass in stars after $\sim$0.8/1.6/5.3 Gyr: had we chosen the mass in stars at redshift of observation as a definition of stellar mass, such a galaxy would not be included in our sample at high redshift, then would enter it at a given epoch and drift from one mass bin to another at an irregular pace, thus biasing the result.
We discuss in Section \ref{sec:massdef} the impact of this definition on our results.

% FIGURE: MASS DEFINITION

We used four galaxy mass bins
($\log (M/M_{\sun})$ in [9.92, 10.33[, [10.33, 10.73[, [10.73, 11.13[, and [11.13, +$\infty$[).
We hereafter refer to those galaxy mass bins by using the mean value of each bin (using a \citealt{schechter76} function), i.e. $\langle \log (M/M_{\sun})  \rangle \sim 10.14$, 10.54, 10.94, and 11.47.
These (model) masses are computed for solar metallicity, a formation redshift of $z_{form}=5$ (setting $z_{form}$ to 4 or 6 does not change our results), and either SSP or an exponentially declining star-forming $\tau$ model with $0 < $ SFH $\tau$ (Gyr) $\le 10$.
The mass values used to define mass bins correspond to the mass of an SSP model having at $z=0$ a $V$-band absolute rest-frame magnitude $M_V$ of -17.8,-18.8,-19.8, and -20.8.

The depth of the data limits the sampled mass range: for each cluster, we required that the lowest mass cut is brighter than a signal-to-noise ratio of 5 in the considered color-magnitude diagram.
This constraint -- probing higher masses at higher redshifts -- restricts the mass range probed, but ensures that we work on mass-complete samples and allows a direct comparison of $f_{blue}$ values at different redshift and mass bins.

%******************************
% BLUE/RED DEFINITION
%******************************
\subsubsection{Blue/red definition}

\begin{figure}
\includegraphics[width=\linewidth]{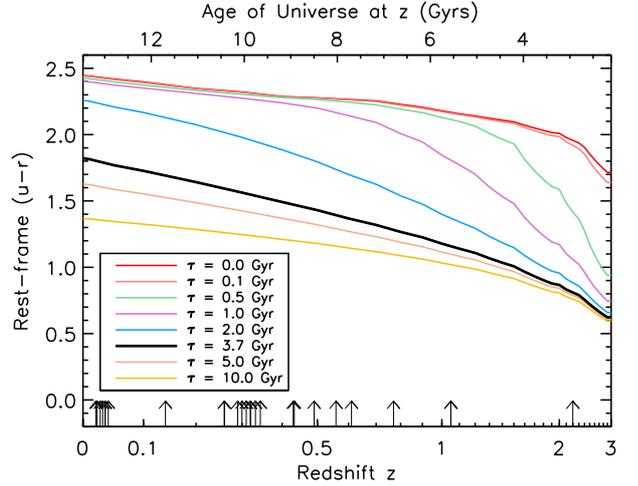}
\caption{Model rest-frame $u-r$ color evolution for different exponentially declining $\tau$-SFHs (CB07 models, solar metallicity and $z_{form}=5$).
According to our definition a galaxy is classified as red (resp. blue) if it is redder (resp. bluer) than a model with $\tau = 3.7$ Gyr.
The vertical arrows indicate the redshifts of our cluster sample.}
\label{fig:taucol_evol}
\end{figure}

We define a galaxy as blue if it is bluer than a CB07 model with $\tau=3.7$ Gyr, as in \citet{andreon04,andreon06,andreon08a}, \citet{loh08}, and \citet{raichoor12}.
This galaxy will be bluer by 0.2 mag in $B-V$ than red-sequence galaxies at $z = 0$ \citep[which would be a blue galaxy by the original definition of][]{butcher84}.
As a comparison, our definition agrees at $z = 0$  with that used in \citet{peng10}, and efficiently splits the red-sequence and the blue cloud of their SDSS data.
The rationale behind our choice is to take into account the stellar evolution of galaxies with time.
Figure \ref{fig:taucol_evol} illustrates the rest-frame $u-r$ color evolution for different exponentially declining $\tau$-SFHs: the different tracks do not cross each other.
In other words, if a galaxy follows an exponentially $\tau$-SFH with $\tau < 3.7$ Gyr (resp. $\tau > 3.7$ Gyr) during its lifetime, its color according to our definition will remain red (resp. blue).
However, we note that this figure assumes that all galaxies have the same $z_{form}$.
We will return to this point in Sect. \ref{sec:conclusion}.

In addition, we stress that the independence of $f_{blue}$ on the probed rest-frame color (Sect. \ref{sec:color}) is a consequence of our choice to define red and blue galaxies (see \citealt{fairley02} for an opposite result when using the red-sequence to split red and blue galaxies).\\

% FIGURE: COL. VS. TAU EVOLUTION

%@@@@@@@@@@@@@@@@@@@@@@@@@@@@@@@@@@@@@@@@@@@
% RESULTS
%@@@@@@@@@@@@@@@@@@@@@@@@@@@@@@@@@@@@@@@@@@@

\section{Results \label{sec:results}}

We present in this section the dependence of $f_{blue}$ with galaxy mass, cluster-centric distance and redshift.
We recall that the evolution of $f_{blue}$ with redshift traces the fraction of cluster galaxies that have stopped forming stars, under the assumption of an average exponentially declining star formation history for cluster galaxies.
Indeed, under our working assumptions and our definition of blue/red galaxies, any quenching implies a lowering of $f_{blue}$ with redshift, whereas no quenching implies an $f_{blue}$ constant with redshift.
Hence our results present the evolution with redshift of the star formation activity in clusters as a function of galaxy mass and environment.

Before performing a general analysis, we start by a simple analysis that makes partial use of the data (Sect. \ref{sec:BOE_indiv}) and that therefore does not show trends in a similar clear way (Sects. \ref{sec:BOE_stack} \& \ref{sec:BOE_model}).
Our main result is presented in Section \ref{sec:BOE_model} and illustrated in Figure \ref{fig:fblue_dep}: environmental and mass quenching are separable, environmental quenching does not change with epoch, and mass quenching is a dynamical process, i.e. its  evolutionary rate is mass-dependent.

%------------------------------------------------------------------------------------------------------------------------
% RESULTS FOR INDIVIDUAL CLUSTERS
%------------------------------------------------------------------------------------------------------------------------
\subsection{Results for individual clusters \label{sec:BOE_indiv}}

For each cluster, galaxy mass bin, and radial area we computed the blue fraction $f_{blue}$ that accounts for residual background galaxies (i.e. along the line of sight, kept by the criteria of Eqs.(\ref{eq:zphot_crit1}) \& (\ref{eq:zphot_crit2}), and not belonging to the cluster) using our control samples (as defined in Sect. \ref{sec:bkg_rmv}) and following the Bayesian methods introduced in \citet{andreon06}.
We adopt uniform priors for the parameters.

\begin{figure*}
\includegraphics[width=\linewidth]{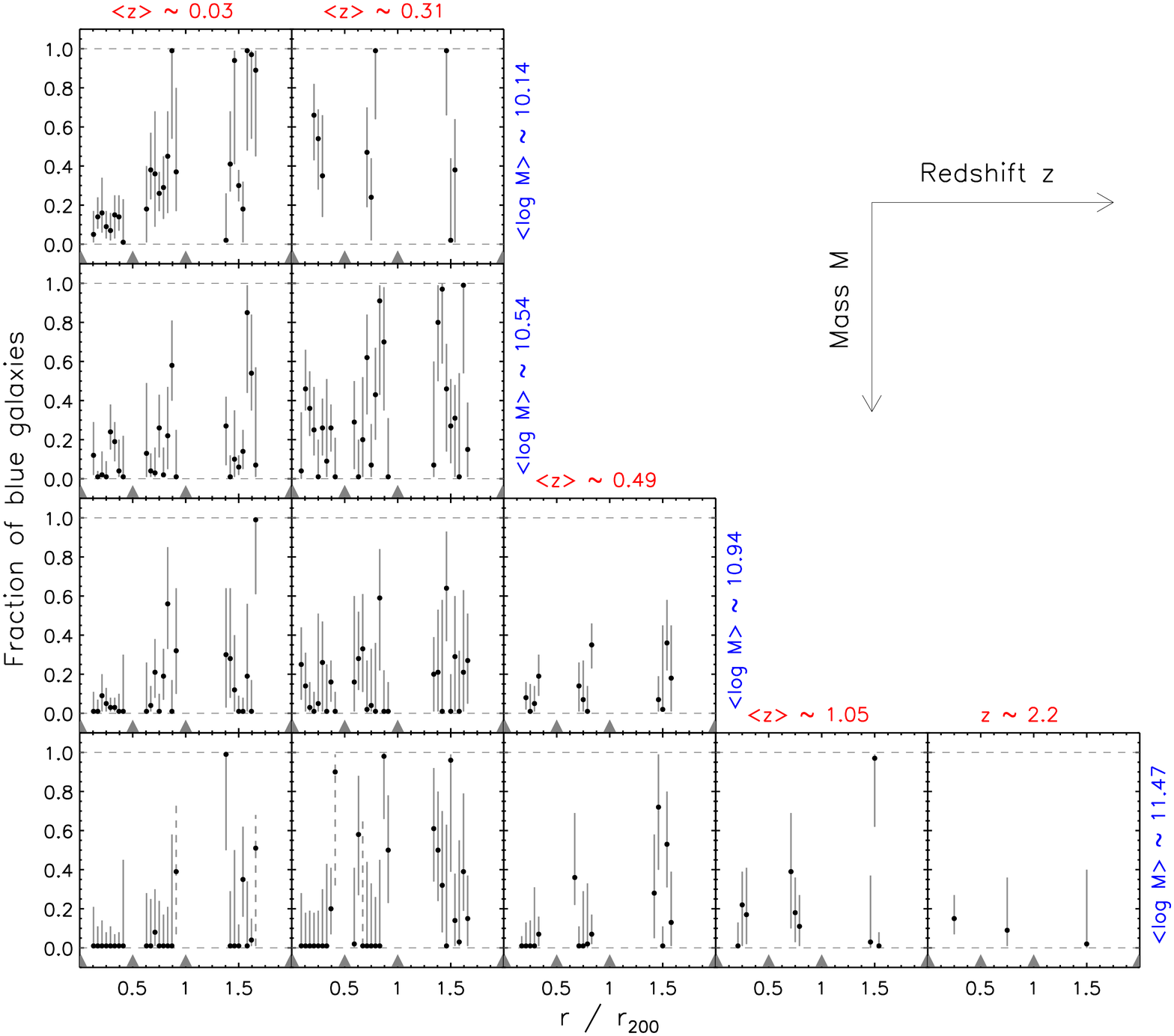}
\caption{
$f_{blue}$ for individual clusters as a function of cluster-centric distance ($r/r_{200}$) for different bins of redshift (increasing rightward) and galaxy mass (increasing downward).
The computed value of $f_{blue}$ already takes into account the mean aging of stars with increasing redshift by defining the color compared to an exponentially declining SFH model with $\tau = 3.7$ Gyr.
Error bars represent the shortest interval including 68\% of the possible posterior values for $f_{blue}$, and are plotted as a dashed line when this interval is larger than 0.66, indicating that $f_{blue}$ is not constrained at all.
Radial cluster-centric bins are indicated by gray filled triangles on the x-axis and horizontal gray dashed lines indicate the minimum and maximum allowed values for $f_{blue}$.
}
\label{fig:BOE_indiv}
\end{figure*}

Figure \ref{fig:BOE_indiv} displays the individual $f_{blue}$ profiles for our cluster sample, grouped by redshift and mass bins.
We defined our redshift bins so that each bin spans a period of $\sim$2 Gyr ($0 \le z < 0.16$, $0.16 \le z < 0.37$, $0.37 \le z < 0.65$, $0.65 \le z < 1.1$, and $z \sim 2.2$).
We hereafter refer to those redshift bins by using the median value of our cluster sample in each bin, i.e. $\langle z \rangle \sim 0.03$, 0.31, 0.49, 1.05, and 2.2.
We can already observe some indications of the trends that we later perceive in stacking analysis.

We observe a different behavior for different galaxy mass bins: at $\langle z \rangle \sim  0.31$, $f_{blue} (r/r_{200} \le 0.5)$ decreases (from $\sim$0.4 to $\sim$0) with increasing galaxy mass.
$f_{blue}$ seems to increase with increasing $r/r_{200}$: for instance, at $\langle z \rangle \sim 0.03$ for the less massive galaxies, $f_{blue}$ increases with increasing cluster-centric distance.
For the most massive galaxies, $f_{blue} (r/r_{200} \le 0.5) \sim 0$ at all redshifts.

However, the unavoidable noise -- because of the finite and usually small number of member galaxies per cluster at a given mass and cluster-centric distance bins -- makes it difficult to draw a general picture.

% FIGURE: BOE INDIV

%------------------------------------------------------------------------------------------------------------------------
% RESULTS FOR STACKED CLUSTERS
%------------------------------------------------------------------------------------------------------------------------
\subsection{Results for stacked clusters \label{sec:BOE_stack}}

\begin{figure*}
\includegraphics[width=\linewidth]{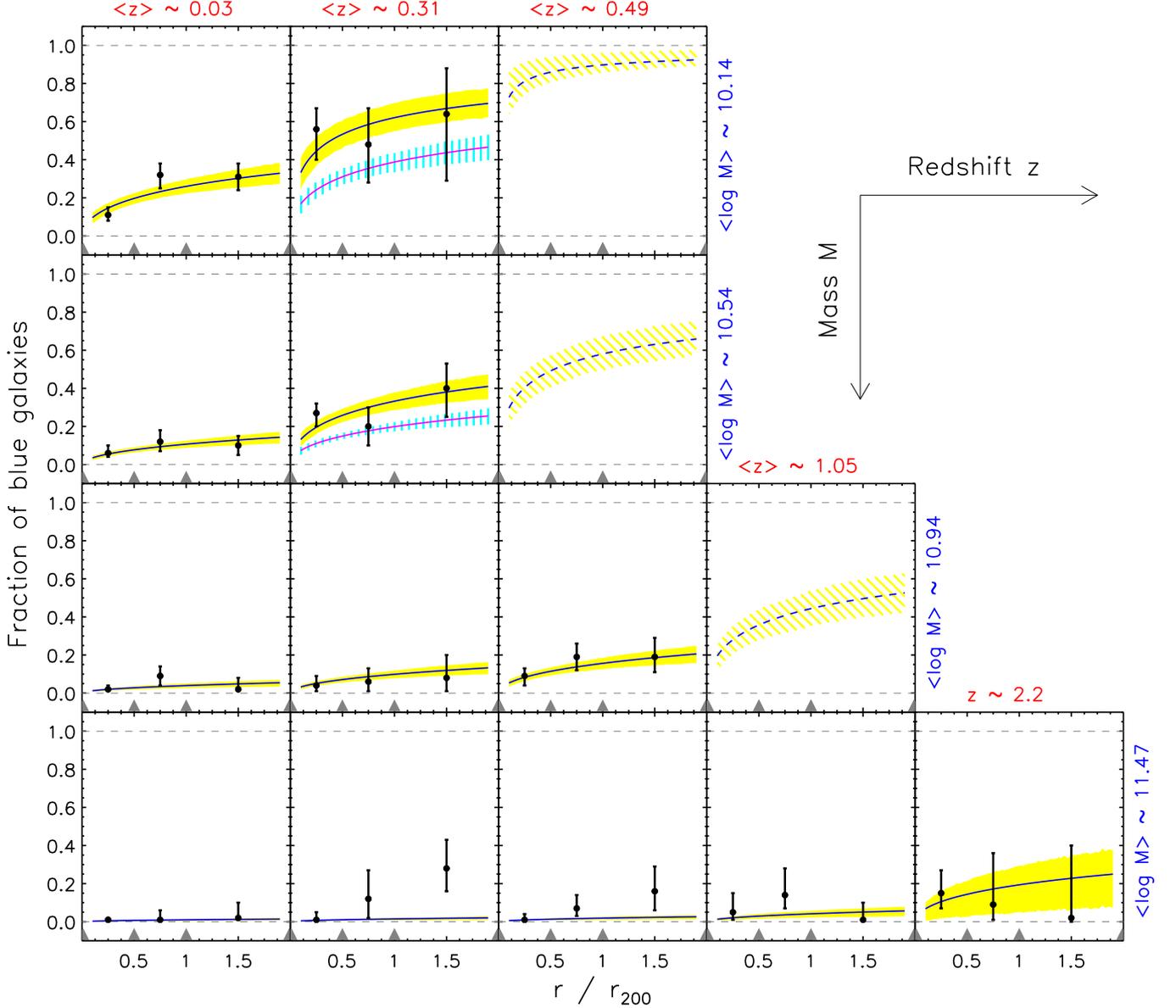}
\caption{
$f_{blue}$ for stacked clusters as a function of cluster-centric distance ($r/r_{200}$) for different bins of redshift (increasing rightward) and galaxy mass (increasing downward).
The computed value of $f_{blue}$ already takes into account the mean aging of stars with increasing redshift by defining the color compared to an exponentially declining SFH model with $\tau = 3.7$ Gyr.
Error bars represent the shortest interval including 68\% of the possible posterior values for $f_{blue}$.
Radial cluster-centric bins are indicated by gray filled triangles on the x-axis and horizontal gray dashed lines indicate the minimum and maximum allowed values for $f_{blue}$.
Yellow shaded areas with a solid blue line represent the posterior mean and  68\% confidence interval of the modeling of Eq.(\ref{eq:model}), fitting the 225 individual $f_{blue}$ measurements.
Yellow slanting hatched areas with a dashed blue line represent the prediction/extrapolation of this model for bins where we do not have data.
Cyan vertically hatched areas with solid magenta line (in two panels only) illustrate how the model fit fails if the $\zeta$ term in Eq. \ref{eq:model} is not included.
}
\label{fig:BOE_stack}
\end{figure*}

We now stacked our sample according to redshift and mass bins to strengthen the estimated $f_{blue}$.
For each panel of Figure \ref{fig:BOE_indiv}, we computed the mean $f_{blue}$ for all clusters belonging to this panel, multiplying the individual data likelihoods, and then using the Bayes theorem as in the previous section.
We recall that the combined data are independent, in particular the control samples.
The values of $f_{blue}$ are reported in Table \ref{tab:fblue_stack} and displayed in Figure \ref{fig:BOE_stack} as dots with error bars.
Now the trends are emerging quite clearly.

% FIGURE: BOE STACK

% R/R200 DEPENDENCE
Regarding the dependence of $f_{blue}$ with cluster-centric distance, our data seem to indicate an increase in $f_{blue}$ with increasing $r/r_{200}$, for most redshift and mass bins.
The cluster color profile at given galaxy mass tends to flatten with decreasing redshift.
In particular, the flat profile is achieved at the lowest redshift and most massive galaxies.
% Z DEPENDENCE
For all mass and radial bins, our data suggest a decrease of $f_{blue}$ with decreasing redshift.
For our two lowest mass bins, this evolutionary trend is strong.
For the highest mass bin, this decrease is marginal, because $f_{blue}$ already has very low values at high redshifts.

% M DEPENDENCE
Finally, we also see a trend when looking at the dependence of $f_{blue}$ with galaxy mass.
Indeed, for our two lowest redshift bins, where our data span four mass bins, we observe a clear trend at all radial bins: at fixed redshift and cluster-centric distance, $f_{blue}$ decreases with increasing galaxy mass.
For instance, for the innermost radial bin and $\langle z \rangle \sim 0.03$, $f_{blue}$ decreases continuously (with increasing galaxy mass) from $0.11_{-0.03}^{+0.04}$ to $0.01_{-0.01}^{+0.01}$.
For $\langle z \rangle \sim 0.49$, though our data only span two mass bins, we observe the same trend.
Our analysis shows that \textit{there is -- in addition to the known mean aging of the stellar populations -- a decrease at all radii in $f_{blue}$ with decreasing redshift, the intensity of which decreases when the galaxy mass increases, and disappears for the most massive galaxies}.

%MISFITS...
We note that in one panel of Figure \ref{fig:BOE_stack} ($\langle z \rangle \sim 0.31$ and $\langle \log (M/M_{\sun}) \rangle \sim 11.47$), our data present an odd behavior for $r/r_{200} \ge 1$: the value of $f_{blue}$ is quite high and does not reflect the general trend, as the mismatch with the fit (detailed below) illustrates.
However, we keep in mind that the 0 value is just at 2$\sigma$.

% TABLE: FBLUE FOR STACKED SAMPLES
\begin{table*}
\centering
\caption{$f_{blue}$ for our cluster sample stacked by redshift and galaxy mass bins. \label{tab:fblue_stack}}
\begin{tabular}{l l l l l l l}
\hline
\hline
Mass bin        & Radial bin    &       $0 \le z < 0.16$        &        $0.16 \le z < 0.37$    &       $0.37 \le z < 0.65$     &       $0.65 \le z < 1.10$     &       $z \sim 2.2$            \\
($\log M/M_{\sun}$)&($r/r_{200}$)&&&&\\
\hline
$[9.92,10.33[$  &       [0, 0.5[                &       0.11$_{-0.03}^{+0.04}$  &       0.56$_{-0.16}^{+0.11}$  &       -       &       -       &       -       \\
        &       [0.5, 1[                &       0.32$_{-0.07}^{+0.06}$  &       0.48$_{-0.20}^{+0.19}$  &       -       &       -       &       -       \\
        &       [1\phantom{.5}, 2[              &       0.31$_{-0.07}^{+0.07}$  &       0.64$_{-0.35}^{+0.24}$  &       -       &       -       &       -       \\
&&&&&&\\
$[10.33,10.73[$ &       [0, 0.5[                &       0.06$_{-0.02}^{+0.04}$  &       0.27$_{-0.07}^{+0.05}$  &       -       &       -       &       -       \\
        &       [0.5, 1[                &       0.12$_{-0.05}^{+0.06}$  &       0.20$_{-0.10}^{+0.10}$  &       -       &       -       &       -       \\
        &       [1\phantom{.5}, 2[              &       0.10$_{-0.05}^{+0.05}$  &       0.40$_{-0.15}^{+0.13}$  &       -       &       -       &       -       \\
&&&&&&\\
$[10.73,11.13[$ &       [0, 0.5[                &       0.02$_{-0.01}^{+0.02}$  &       0.04$_{-0.03}^{+0.05}$  &       0.09$_{-0.05}^{+0.04}$  &       -       &       -       \\
        &       [0.5, 1[                &       0.09$_{-0.05}^{+0.05}$  &       0.06$_{-0.05}^{+0.07}$  &       0.19$_{-0.07}^{+0.07}$  &       -       &       -       \\
        &       [1\phantom{.5}, 2[              &       0.02$_{-0.01}^{+0.06}$  &       0.08$_{-0.07}^{+0.12}$  &       0.19$_{-0.08}^{+0.10}$  &       -       &       -       \\
&&&&&&\\
$11.13 \le$     &       [0, 0.5[                &       0.01$_{-0.01}^{+0.01}$  &       0.01$_{-0.01}^{+0.04}$  &       0.01$_{-0.01}^{+0.03}$  &       0.05$_{-0.04}^{+0.10}$  &       0.15$_{-0.08}^{+0.12}$  \\
        &       [0.5, 1[                &       0.01$_{-0.01}^{+0.05}$  &       0.12$_{-0.10}^{+0.15}$  &       0.07$_{-0.04}^{+0.07}$  &       0.14$_{-0.07}^{+0.14}$  &       0.09$_{-0.08}^{+0.27}$  \\
        &       [1\phantom{.5}, 2[              &       0.02$_{-0.01}^{+0.08}$  &       0.28$_{-0.12}^{+0.15}$  &       0.16$_{-0.10}^{+0.13}$  &       0.01$_{-0.00}^{+0.09}$  &       0.02$_{-0.02}^{+0.38}$  \\
&&&&&&\\
\hline
\end{tabular}
\tablefoot{The values of $f_{blue}$ are those plotted in Figure \ref{fig:BOE_stack}.}
\end{table*}

%------------------------------------------------------------------------------------------------------------------------
% RESULTS FOR THE WHOLE SAMPLE
%------------------------------------------------------------------------------------------------------------------------
\subsection{Results for the whole sample \label{sec:BOE_model}}

We now fitted the 225 values of $f_{blue}$ (26 clusters, 3 radial bins, 1 to 4 galaxy mass bins) at once by modeling at the same time the dependence of $f_{blue}$ on galaxy mass $M$, redshift $z$ and cluster-centric distance $r/r_{200}$.
In this section, we are no longer ignoring redshift differences inside each panel, and the collective use of the whole dataset allows us to strengthen more the trends even.

As detailed below, to described the data, we also need an interaction term between galaxy mass and redshift, a term that allows galaxies of different mass to evolve at different rates.
We use the following model:
\begin{eqnarray}
{f_{blue}} \left(r/r_{200},M,z \right) &=& \texttt{ilogit} \, \Bigl[ A_0 + \nonumber\\
&&	\alpha \cdot \log \left(r/(0.25 \cdot r_{200}) \right) + \nonumber \\
&&	\beta \cdot (\log(M/M_{\sun}) - 11) + \nonumber\\
&&	\gamma \cdot (z-0.3) + \nonumber \\
&&	\zeta \cdot(\log(M/M_{\sun})-11) \cdot (z-0.3) \Bigr],
\label{eq:model}
\end{eqnarray}
where $\texttt{ilogit}(x) = (1+\exp(-x))^{-1}$ ensures that $0 \le f_{blue} \le 1$.
We adopt uniform priors for the parameters $A_0$, $\alpha$, $\beta$, $\gamma$, and $\zeta$.

The model fit results are plotted in Figure \ref{fig:BOE_stack} as solid blue lines and yellow shaded areas (68\% confidence interval), while the model prediction/extrapolation is plotted as dashed blue lines and yellows hatched areas.

% SIMPLEST MODEL
The only motivation behind the chosen parametrization is the adoption of an additive model that fits our data and ensures that $0 \le f_{blue} \le 1$.
The adopted model is the simplest one fulfilling those criteria.
If we do not include the $\zeta$ term, the model cannot fit data in the redshift bins $\langle z  \rangle \sim 0.31$ for galaxies within our two lowest mass bins (see cyan vertically hatched areas in Figure \ref{fig:BOE_stack}) and, to a lesser extent, in the panel $\langle z  \rangle \sim 0.49$ and $\langle  \log(M/M_{\sun}) \rangle \sim 10.94$.
In a similar manner, if we replace the term crossing $M$ and $z$ by a term crossing $r$ and $z$ or $r$ and $M$, the model fails in fitting the data.
Furthermore, we note that adding a term crossing $r$ and $z$ (in addition to our $\zeta$ term) does not improve the quality of the fit.
\textit{The $f_{blue}$ evolution with redshift is therefore more sensitive to the galaxy mass than to its cluster-centric distance}.

% FIGURE: JAGS SUPERSTACK PARAMS
\begin{figure*}
\includegraphics[width=\linewidth]{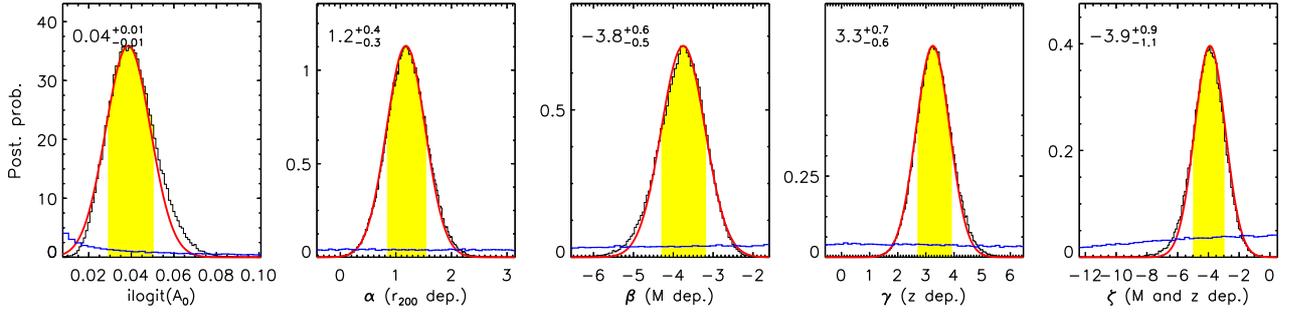}
\caption{
For each parameter fitted with the model of Eq.(\ref{eq:model}), we plot as a black lines the posterior probability function and as yellow shaded area the shortest interval including 68\% of the possible posterior values.
We report the point estimate and the 68\% confidence interval on each panel, and overplot as a thick red line the corresponding Gaussian curve.
The blue lines represent the uniform prior used for the fit.
}
\label{fig:superstack_params}
\end{figure*}

% PARAMETER POSTERIORS
The posterior probabilities for the model parameters are plotted in Figure \ref{fig:superstack_params}, which shows how our knowledge about the parameters changes from before (prior, blue line) to after (posterior, black/red lines) the data.
In short, the posterior distribution of all parameters is much more concentrated than the prior, i.e. data are highly informative about these parameters and conclusions on these parameters do not depend on the adopted prior.

% PARAMETERS SHORT ANALYSIS
The values agree with our qualitative trends put forth in Sect. \ref{sec:BOE_stack}, with $f_{blue}$ increasing with cluster-centric distance ($\alpha = 1.2_{-0.3}^{+0.4}$), decreasing with galaxy mass ($\beta = -3.8_{-0.5}^{+0.6}$), and increasing with redshift ($\gamma = 3.2_{-0.5}^{+0.7}$).
$\zeta$ measures the relative speed -- when compared to the evolution of galaxies with $\log (M/M_{\sun}) = 11$ -- at which $f_{blue}$ evolves as a function of galaxy mass.
Because $\zeta$ is negative ($\zeta = -4.1_{-0.9}^{+1.1}$), galaxies with lower masses will on average evolve on longer timescales.
We note that our model is constrained at $z \gg 1$ only by the JKCS\,041 cluster (if we repeat the fit without the JKCS\,041 data, the model appears to be entirely unconstrained in the $\langle \log(M/M_{\sun}) \rangle \sim 11.47$ and $z \sim 2.2$ window).

\begin{figure*}
\includegraphics[width=\linewidth]{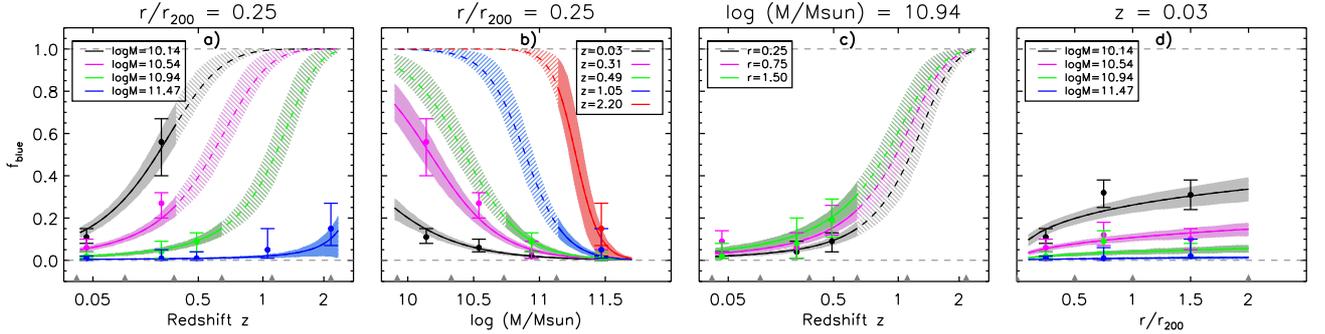}
\caption{
Dependence of $f_{blue}$ on $M$ and $z$ at fixed $r/r_{200}$ (panels a and b), on $z$ and $r/r_{200}$ at fixed $M$ (panel c), and on $M$ and $r/r_{200}$ at fixed $z$ (panel d).
Data points with error bars are the same as in Figure \ref{fig:BOE_stack}.
Shaded areas represent the posterior mean and  68\% confidence interval of the modeling of Eq.(\ref{eq:model}), fitting the 225 individual $f_{blue}$ measurements.
Prediction/extrapolation of this model for bins where we do not have data are plotted as hatched areas and dashed lines.
X-axis bins are indicated by gray filled triangles on the x-axis and horizontal gray dashed lines indicate the minimum and maximum allowed values for $f_{blue}$.
}
\label{fig:fblue_dep}
\end{figure*}

% OTHER PLOT
To better visualize the dependence of $f_{blue}$ on those three parameters ($M$, $r/r_{200}$, and $z$), we display in Figure \ref{fig:fblue_dep} the variations of $f_{blue}$ on two of them when fixing the third one.
The information (data and model fit/predictions) is exactly the same as in Figure \ref{fig:BOE_stack}, only presented differently.
We recall that our model is only constrained where there are some data (shaded areas), and that the fit was performed on the individual 225 values of $f_{blue}$.
This figure illustrates on the one hand that at fixed cluster-centric distance, $f_{blue}$ evolves at different paces for different galaxy stellar masses (panels a and b), captured by the $\zeta$ term in our model.
On the other hand, we can see in panel c that the model evolution of $f_{blue}$ with $z$ at fixed galaxy mass has a similar shape for our values of $r/r_{200}$.
Lastly, panel d summarizes the radial profiles of $f_{blue}$ at a fixed redshift for different galaxy mass bins.

% FIGURE: MODEL: FBLUE VARIOUS DEPENDENCES

%------------------------------------------------------------------------------------------------------------------------
% BOE FOR Mv<-19.3
%------------------------------------------------------------------------------------------------------------------------
\subsection{Results for the original Butcher \& Oemler cut}
\begin{figure*}
\includegraphics[width=\linewidth]{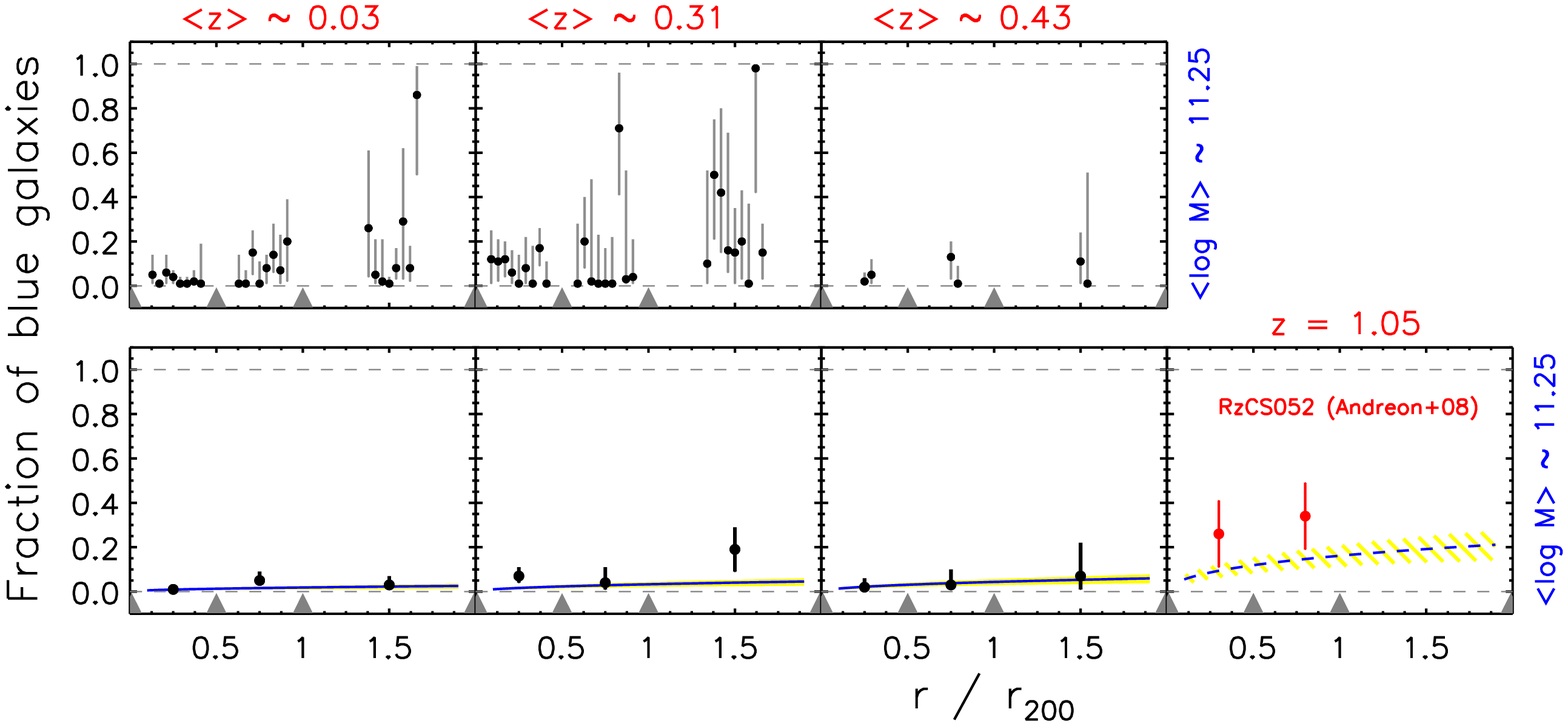}
\caption{
Evolution of $f_{blue}$ for individual (top panels) and stacked (bottom panels) clusters as a function of cluster-centric distance ($r/r_{200}$) for different bins of redshift (increasing rightward) for a mass-selected sample with a mass corresponding to the original $M_V$ in \citet{butcher84}.
Symbols are as in Figures \ref{fig:BOE_indiv} and \ref{fig:BOE_stack}.
We add a new panel at $z=1.05$ for model prediction, where we overplot the measured values of $f_{blue}$ for the RzCS\,052 cluster in \citet{andreon08a}.
}
\label{fig:BOE_-19.3}
\end{figure*}

Adopting a mass threshold corresponding to the original \citet{butcher84} one ($\log(M/M_{\sun}) \ge 10.53$), we observe no evolution of $f_{blue}$ with redshift for our data ($z \le 0.43$) (see Figure \ref{fig:BOE_-19.3}).
However, when extrapolating the model to $z = 1.05$, we observe an increase in $f_{blue}$, in broad agreement with the measurement done by \citet{andreon08a} for the RzCS\,052 cluster.

That almost no evolution is found at $z \le 0.43$, whereas we do detect one for $\langle \log (M/M_{\sun}) \rangle \sim 10.54$, nicely illustrates the benefit of splitting our sample into fine galaxy mass bins.

% FIGURE: BOE -19.3

%@@@@@@@@@@@@@@@@@@@@@@@@@@@@@@@@@@@@@@@@@@@
% DISCUSSION/CONCLUSION
%@@@@@@@@@@@@@@@@@@@@@@@@@@@@@@@@@@@@@@@@@@@

\section{Discussion and conclusion \label{sec:conclusion}}

Our aim was to put constraints on the different processes responsible for the cessation of star formation activity ("quenching") in clusters through a careful study of the dependence of the fraction of blue galaxies $f_{blue}$ in clusters that span a broad redshift baseline.
We have built a cluster sample consisting of seven clusters ($0.02 \le z_{spec} \le 0.04$) from the HIFLUGCS survey, eighteen clusters ($0.14 \le z_{spec} \le 1.05$) from the XMMLSS survey, and JKCS\,041 ($z \sim 2.2$).
Because the HIFLUGCS and XMMLSS surveys are X-ray selected, our cluster sample is unbiased regarding the fraction of blue galaxies $f_{blue}$ at a given cluster mass; in addition, our clusters were chosen to have similar masses.
For different galaxy mass and cluster-centric distance bins, we carefully estimated $f_{blue}$ for each cluster and fitted the dependence of $f_{blue}$ with redshift, galaxy mass and cluster-centric distance.
We did not attempt to invert the noisy data because of the degeneracy between $z_{form}$, age (and metallicity), but instead we modified the model until it fitted the data.
The reference model has an exponentially declinining SFH, and hence an evolving mass (if stars are formed, mass in stars cannot be fixed).
Deviations from this reference model highlight the evolution and respective role of the quenching in clusters due to galaxy mass or to environment.

Our main result is displayed in Figures \ref{fig:BOE_stack} and \ref{fig:fblue_dep}.
We recall that in those figures we have already taken into account that galaxies are on average younger at higher redshifts, by using an exponentially declining $\tau$-model to define whether a galaxy is red or blue.
This means that any measured evolution of $f_{blue}$ with redshift reflects a phenomenon in addition to the younger mean age of the Universe and the secular increase in the star formation rate.
We also recall that our model fitting was performed on the 225 individual measurements of $f_{blue}$.
Moreover, our definition of galaxy mass ensures that, for each mass bin, the selected high-redshift galaxies are those that would end up in the low-redshift galaxies selection.

% GENERAL TRENDS
We are therefore able to disentangle for the first time the role of secular evolution, galaxy mass and cluster-centric distance on galaxy evolution in clusters.
We found that $f_{blue}$ decreases with decreasing redshift $z$, with decreasing cluster-centric distance $r/r_{200}$, and with increasing galaxy mass $M$.
This means that the processes responsible for the cessation of star formation in clusters are effective at all epochs ($ z \lesssim 2.2$) and more effective in denser environments and for more massive galaxies.
Our (simple) modeling of the data shows that the dependence of $f_{blue}$ on galaxy mass evolves with redshift, while this is not the case for the dependence of $f_{blue}$ on cluster-centric distance: the intensity of mass quenching evolves with redshift at different paces for different galaxy masses, whereas the environmental quenching does not evolve with redshift.
Additionally, our data did not need any correlation between galaxy mass and cluster-centric distance in the model fitting, which means that mass quenching and environmental quenching are separable.

%------------------------------------------------------------------------------------------------------------------------
% COMPARISON WITH PRESOTTO-LIKE DEFINITION OF BLUE/RED
%------------------------------------------------------------------------------------------------------------------------
\subsection{Remark on taking into account secular evolution when estimating $f_{blue}$}
We illustrate in this section the advantage of taking into account secular evolution when estimating $f_{blue}$: using a non-evolving color threshold \citep[e.g.][]{presotto12} will result in using a redder color to split blue and red galaxies, hence in increasing $f_{blue}$.

As a test, we repeated our analysis, with a non-evolving, mass-dependent definition of blue/red galaxy.
For low redshifts ($z \lesssim 0.5$), this leads to results marginally different from ours, because the colors used in the two methods to split blue and red galaxies do not differ significantly.
However, for higher redshifts, we do observe a notable difference between the two methods, because the difference between the two definitions of blue/red galaxies will increase with increasing redshift.
For instance, using a non-evolving color threshold leads to larger blue fractions at $z \gtrsim 1$ ($0.2 \lesssim f_{blue} \lesssim 0.6$) for stacked data.
Since $f_{blue}$ depends on $r/r_{200}$, a non-evolving color threshold also tilts the $f_{blue}$ profile.
Overall, the results using a non-evolving color threshold are much harder to interpret, because there is an additional effect that depends on the redshift (and potentially the radial) bin(s) under inspection.

%------------------------------------------------------------------------------------------------------------------------
% MASS=MASS IN STARS AT ZOBS
%------------------------------------------------------------------------------------------------------------------------
\subsection{Remark on using evolved mass definition \label{sec:massdef}}

We remark that using the \textit{mass in stars at the redshift of observation}, $M_{zobs}$, as a mass definition does not appreciably change our conclusions: the estimated values of $f_{blue}$ are slightly lower, but the trivariate dependence of $f_{blue}$ on redshift, galaxy mass and cluster-centric distance remains similar (the $\alpha$, $\beta$, $\gamma$, and $\delta$ parameters vary by less than 1$\sigma$).

Indeed, as illustrated in Figure \ref{fig:massdef}, lines of constant mass at the redshift of observation in the color-magnitude diagrams are similar to those used in the study, but for blue colors.
For a considered mass bin, using $M_{zobs}$ instead of $M$ will remove galaxies in the light-gray shaded area, but add those in the dark-gray shaded area.
Because of the mass function shape, the former will on average be slightly more numerous than the latter: the net effect will be to slightly decrease the number of blue galaxies, hence $f_{blue}$.

%------------------------------------------------------------------------------------------------------------------------
% M AND Z
%------------------------------------------------------------------------------------------------------------------------
\subsection{Dependence of $f_{blue}$ on $M$ and $z$}
The data and their fit point out two results:
\begin{itemize}
\item The average SFH is more complex than exponentially declining SFHs, as quantified by the $\gamma$ and $\zeta$ terms.
Indeed, if cluster galaxies had on average exponentially declining SFHs, we should not observe any change of the $f_{blue}$ radial profile with redshift (at fixed galaxy mass).
Low-mass galaxies may have a briefly enhanced SFH because of starbust, hence temporarily a color bluer than an SFH $\tau = 3.7$ Gyr model, after which their color turns redder than an SFH $\tau = 3.7$ Gyr model.
For our lowest mass bin, this mechanism would still be at work at $z < 0.37$.
We also considered the case of a later $z_{form}$ for low-mass galaxies.
We experimented with some possible choices without being able to remove the observed evolution of $f_{blue}$ with redshift, although our tests are not exhaustive and we are reaching the limit of the data.
\item There is a differential evolution of $f_{blue}$ with the galaxy mass, i.e. mass quenching is a dynamical process. Galaxies with smaller masses evolve later: for instance, at $z \le 0.37$, $f_{blue}$ for galaxies in our lowest mass bin still evolve significantly, which is not the case for $f_{blue}$ for more massive galaxies.
\end{itemize}

Furthermore, less massive galaxies seem to evolve on a longer timescale than more massive galaxies.
In the $\sim$3 Gyr between our two lowest redshift bins, $f_{blue}$ changes by $\sim$0.4 for low-mass galaxies.
In less time ($\sim$2 Gyr), $f_{blue}$ should change by twice as much at $z\gtrsim2$ for high-mass galaxies.

%------------------------------------------------------------------------------------------------------------------------
% R/R200
%------------------------------------------------------------------------------------------------------------------------
\subsection{Dependence of $f_{blue}$ on $r/r_{200}$}
We now turn to the analysis of the dependence of $f_{blue}$ with cluster-centric distance $r/r_{200}$.
For all galaxy stellar mass and redshift bins, $f_{blue}$ decreases with decreasing $r/r_{200}$, i.e. environmental quenching is effective at all redshifts and stellar masses (Figure \ref{fig:BOE_stack}).
This effect is conspicuous on lower mass galaxies, as $f_{blue}$ is very low for massive galaxies, whatever the cluster-centric distance (one should observe a significant number of clusters at $z \sim 1$-2 to see this effect on massive galaxies).
Previous works clearly established this dependence on $r/r_{200}$ for mass-selected samples \citep[e.g.][]{andreon08a,haines09}, however, our study shows for the first time that this dependence holds for different galaxy mass bins.
In addition, the model fitting our data implies that the mass and environmental quenchings are fully separable, because no crossed term between $M$ and $r/r_{200}$ is required.

Additionally, we found that our data do not require any evolution of this dependence with redshift: our (simple) model requires only a crossed term between $M$ and $z$, none between $r/r_{200}$ and $z$.
This can be interpreted that either this evolution with $z$ of the environmental quenching is not present, or that is of second order when compared to the evolution with $z$ of the mass quenching.
We recall that our cluster sample consists of clusters having similar temperatures at all redshifts, while a self-similarly evolving cluster model \citep{kaiser86} suggests a mild increase with redshift at $0<z<1$.
Nevertheless, current studies tend to show that $f_{blue}$ estimated in the virial radius is independent of cluster mass, at least in the local Universe \citep[e.g.,][]{de-propris04,goto05}.

Finally, we would like to emphasize the importance of the backsplash population on the $f_{blue}$ radial profile and its evolution with redshift: galaxies as far away as two virial radii may have been within the main body of the cluster in the past \citep[e.g.,][]{balogh00,gill05,mahajan11}.

%------------------------------------------------------------------------------------------------------------------------
% CONCLUSION
%------------------------------------------------------------------------------------------------------------------------
\subsection{Conclusion}
% PENG+10
Our study extends the relationship between galaxy mass, star formation rate and environment at $z \lesssim 1$ explored by \citet{peng10} to intermediate-mass cluster environments.
As those authors, we found
that the cessation of star formation ("quenching") due to environment and galaxy mass are separable (we needed no crossed term between $r/r_{200}$ and $M$ to fit our data),
that environmental quenching does not change with epoch (we needed no crossed term between $r/r_{200}$ and $z$ to fit our data),
and that mass quenching is a dynamical process (we did need a crossed term between $M$ and $z$ to fit our data).
The close agreement between the two studies is likely a consequence of the common choices adopted for the two analyses: disentangling galaxy mass and environment, plus taking into account the secular aging of stars with decreasing redshift, as described by \citet{andreon06}.
However, in addition to paying special attention to the statistical aspect of the analysis (which allowed us to provide good confidence intervals on our measurements), our study implemented a finer control on galaxy mass (by considering the mass evolved at $z = 0$) and on environmental estimation (using $r/r_{200}$ is less subject to biases than a density field based on rest-frame $B$-band selection).

% MUZZIN+12
\citet{muzzin12}, investigating a cluster sample at $z \sim 1$ with spectroscopic data, made a similar analysis, studying the role of galaxy mass and environment (but not secular evolution) on star formation.
The availability of spectroscopic data allows estimating the star formation rate, and thus a thorough analysis of this problem.
Regarding the fraction of star-forming galaxies, these authors also concluded that mass and environmental quenching are separable.
Our work extends this study by including the redshift dependence in the analysis, by a better characterization ($r_{200}$) of cluster properties, and by using a cluster sample that is not selected by a galaxy property under study (color).

% DOWNSIZING
Our study extends the \textit{downsizing}-like scenario to cluster environment and all galaxies, which was already established for galaxies with specific properties in the field (e.g., emission line: \citealt{cowie96}; spheroidal: \citealt{treu05}; passive: \citealt{peng10}).
According to this scenario, the properties of the most massive galaxies are established in the very early Universe ($z \gg 1$), while less massive galaxies continue to evolve at redshifts $0 < z < 1$.
We stress that our \textit{downsizing}-like scenario concerns all galaxies, regardless of their morphology or color.
The results of \citet{andreon08a} and \citet{raichoor12} taken as a whole point to a similar conclusion, though for smaller cluster samples.

% CONCLUSION
Our work is the first to study the dependence of $f_{blue}$ in clusters on galaxy mass, redshift, and cluster-centric distance at the same time, with a cluster sample whose properties are well-controlled.
In particular, it clearly demonstrates the need to use galaxy mass as a parameter to better understand the behavior of $f_{blue}$.
This approach is now possible with the wealth of available data.

%@@@@@@@@@@@@@@@@@@@@@@@@@@@@@@@@@@@@@@@@@@@
% BIBLIOGRAPHY
%@@@@@@@@@@@@@@@@@@@@@@@@@@@@@@@@@@@@@@@@@@@

%\bibliography{raichoor_master}

%@@@@@@@@@@@@@@@@@@@@@@@@@@@@@@@@@@@@@@@@@@@
% APPENDIX: R200 ESTIMATION
%@@@@@@@@@@@@@@@@@@@@@@@@@@@@@@@@@@@@@@@@@@@

\appendix

\section{$r_{200}$ estimation \label{app:r200}}

For clusters with $T_X$ measurements \citep{pacaud07,hudson10}, $r_{200}$ is determined using the mass-temperature relation of \citet{finoguenov01}, which has been shown to give robust results for clusters with $T_X \lesssim 4$ keV \citep{willis05}:
\begin{equation}
r_{500} \, \mathrm{(Mpc)} = \frac{0.391 \cdot T_X^{0.63}}{H(z)/70},
\label{eq:r200_Tx}
\end{equation}
where $H (z) = H_0 \sqrt{\Omega_m  (1+z)^3+\Omega_\Lambda}$ describes the redshift evolution of the Hubble parameter in the assumed cosmological model.
Then we use $r_{200} = r_{500} /0.661$ corresponding to a \citet{navarro97} profile with a halo concentration parameter $c=5$.\\

\begin{figure}
\resizebox{\hsize}{!}{\includegraphics{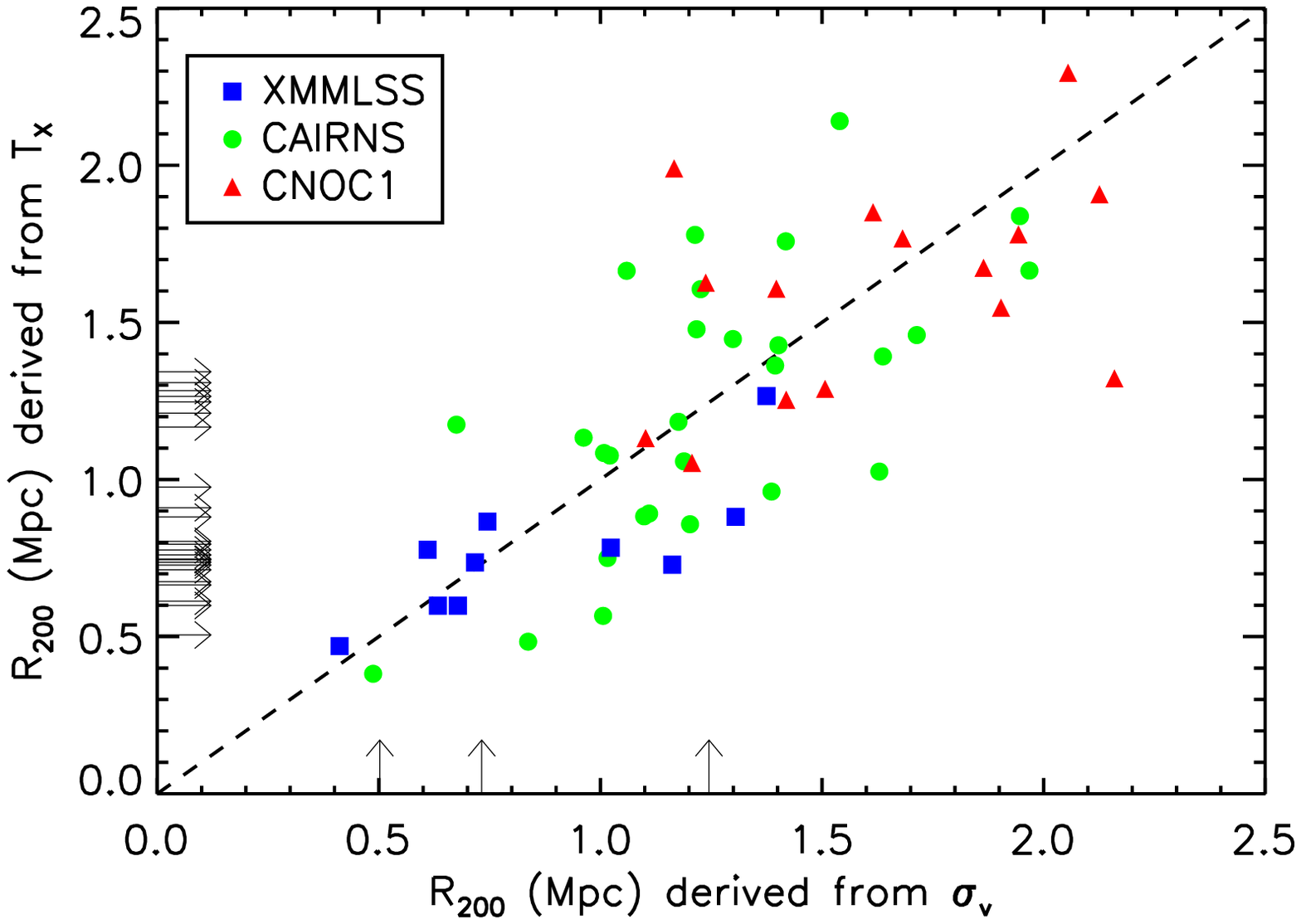}}
\caption{Comparison of $r_{200}$ estimated with Eqs.(\ref{eq:r200_Tx}) \& (\ref{eq:r200_sigma}):
the test sample is made of 10 clusters from the XMMLSS survey \citep{willis05,andreon06,pacaud07}, 28 clusters from the CAIRNS survey \citep{rines06}, and 15 clusters from the CNOC1\citep{muzzin07} and spans $0.003 < z_{spec} < 0.84$, $0.5 < T_X  ($keV$) < 10.3$ and $232 < \sigma_v$ (km s$^{-1}) < 1354$.
There is no bias between the $r_{200}$ estimated from $T_X$ and from $\sigma_v$.
The horizontal (resp. vertical) arrows indicate the $r_{200}$ estimated from $T_X$ (resp. $\sigma_v$) for our cluster sample.
}
\label{fig:r200}
\end{figure}

For clusters with no $T_X$ measurements (XLSSC\,007, XLSSC\,014, and XLSSC\,016), we estimate $r_{200}$ from the cluster velocity dispersion $\sigma_v$, from \citet{willis05}, following the formula
\begin{equation}
r_{200} \, \mathrm{(Mpc)}= \frac{\sqrt{3} \cdot \sigma_v\mathrm{[km s}^{-1}] }{10 \cdot H(z)} \times  0.85 .
\label{eq:r200_sigma}
\end{equation}
This equation is similar to the classical equation for a spherical collapse model \citep[e.g.,][]{carlberg97}, but for the multiplicative coefficient 0.85.
This coefficient comes from a calibration we perform with a compilation of clusters having $T_X$ \textit{and} $\sigma_v$ measurements (see for instance \citealt{proctor11} for a similar systematic difference).
Eq.(\ref{eq:r200_sigma}) ensures that there is no bias between our $r_{200}$ estimated with $T_X$ or $\sigma_v$ (see Figure \ref{fig:r200}).\\

% FIGURE: R200

%@@@@@@@@@@@@@@@@@@@@@@@@@@@@@@@@@@@@@@@@@@@
% APPENDIX: EAZY R-BAND PRIOR
%@@@@@@@@@@@@@@@@@@@@@@@@@@@@@@@@@@@@@@@@@@@

\section{$r$-band prior for photometric redshifts \label{app:r_prior}}

\begin{figure}
\resizebox{\hsize}{!}{\includegraphics{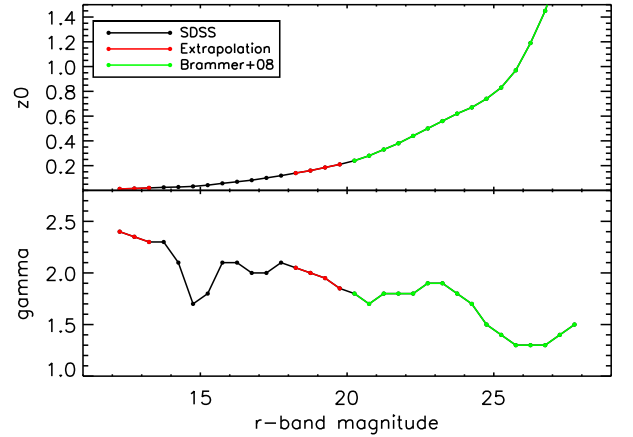}}
\caption{$z_0$ and $\gamma$ parameters used for building $r$-band prior with Eq.(\ref{eq:r_prior}).}
\label{fig:r_prior}
\end{figure}

Because \textsc{Eazy} has been developed for high-$z$ studies, the $r$-band prior does not include magnitudes brighter than 20.
In our SDSS sample, we deal with galaxies as bright as $r \simeq 12$.
We built the prior probabilities $p(z | m_0)$ for $13.5 < r < 18$ as follows.
For each interval of 0.5 mag, we retrieved $\sim$10,000 spectroscopic galaxies from the \texttt{SpecPhoto} view from the SDSS DR8 database and fitted the redshift distribution with a function
\begin{equation}
p(z | m_0) \propto z ^{\gamma} \cdot \exp{[-(z/z_0)^\gamma]},
\label{eq:r_prior}
\end{equation}
as in \citet{brammer08}.
At $13.5 < r$ and $18 < r < 20$, we visually extrapolated $z_0$ and $\gamma$ (cf. Figure \ref{fig:r_prior}), because either there are not enough galaxies or those present in the SDSS are a biased sample (e.g., LRGs, QSOs, etc).

% FIGURE: R-PRIOR

%@@@@@@@@@@@@@@@@@@@@@@@@@@@@@@@@@@@@@@@@@@@
% ACKNOWLEDGMENTS
%@@@@@@@@@@@@@@@@@@@@@@@@@@@@@@@@@@@@@@@@@@@

\begin{acknowledgements}

We thank the anonymous referee for his/her careful reading and suggestions, which improved the clarity of the paper.

We acknowledge financial contribution from the agreement ASI-INAF I/009/10/0 and from Osservatorio Astronomico di Brera.

Based on data from SDSS-III (full text acknowledgement is at \url{http://www.sdss3.org/collaboration/boiler-plate.php}) and
observations obtained with
MegaPrime/MegaCam (\url{http://www.cfht.hawaii.edu/Science/CFHLS/cfhtlspublitext.html}) and 
WIRCAM (\url{http://ftp.cfht.hawaii.edu/Instruments/Imaging/WIRCam/WIRCamAcknowledgment.html})
at CFHT. 

\end{acknowledgements}

\end{document}